# A window view quality assessment framework

Won Hee Ko[1*], Michael G. Kent[2], Stefano Schiavon[1], Brendon Levitt[3,4], Giovanni Betti[1]

[1]*Center for the Built Environment, University of California, Berkeley, CA, USA*

[2]*Berkeley Education Alliance for Research in Singapore (BEARS), Singapore*

[3]*Loisos + Ubbelohde, Alameda, CA, USA*

[4]*California College of the Arts, Oakland, CA, USA*

* Corresponding author:

E-mail address: wonheeko@berkeley.edu

Address: 390 Wurster Hall, Berkeley, CA 94720-1839, USA

# A window view quality assessment framework

The views that windows provide from inside a building affect human health and well-being. Although window view is an important element of architecture, there is no established framework to guide its design. The literature is widely dispersed across different disciplinary fields, and there is a need to coalesce this information into a framework that can be applied into the building design. Based on the literature, we present a framework for what constitutes "view quality." At the basis of our framework, we propose three primary variables: *View Content* (the assessment of visual features seen in the window view); *View Access* (the measure of how much of the view can be seen through the window from the occupant's position); and *View Clarity* (the assessment of how clear the view content appears in the window view when seen by an occupant). Each variable was thematically derived from different sources including daylighting standards, green certification systems, and scientific research studies. We describe the most important characteristics of each variable, and from our review of the literature, we propose a conceptual index that can evaluate the quality of a window view. While discussing the index, we summarize design recommendations for integrating these three variables into the building process and identify knowledge gaps for future research.

## 1. Introduction

One important function of a window is to provide a view to the outside environment (Tregenza and Wilson, 2013). When considering that we spend a significant proportion of our lives indoors (Gifford, 1995; Klepeis *et al.*, 2001), windows are the essential architectural element that allows us to have much-needed contact with outdoors (Bell, 1973; Collins, 1976). The view from a window can have a profound impact on physiology and psychology (Hartig *et al.*, 2003; Kaplan, 1993; Ulrich, 1984) through a range of positive effects on health and well-being (Beute and de Kort, 2014; Küller and Lindsten, 1992), cognitive performance (Jamrozik *et al.*, 2019; Ko *et al.*, 2020), spatial satisfaction (Yildirim *et al.*, 2007), discomfort and stress (Aries *et al.*, 2010; Benfield *et al.*, 2015), and emotion (Ko *et al.*, 2020). Although past review articles (Beute and de Kort, 2014; Boyce *et al.*, 2003; Galasiu and Veitch, 2006; Veitch and Galasiu, 2012; Velarde *et al.*, 2007) have extensively documented the myriad effects of windows, daylight, and views on occupant health and wellbeing, to our knowledge, no review has yet to compile the literature into a comprehensive design framework that can be used by architects or practitioners. To address this issue, we propose a unified framework for view quality to systematically identify the key variables, provide guidance in the design process, and suggest future areas of research.

### *1.1. Problem identification*

View can be defined as: "*What you can see from a particular place*" (Cambridge University Press, 2020). While we use a similar definition in our work, views inside buildings occur when (day)light reflects off outdoor surfaces and this visual information is transmitted through windows (Tregenza and Wilson, 2013). Therefore, window views and (day)light cannot always be considered separate entities. It may be for this reason that some daylighting standards (CEN/TC 169, 2018; SLL, 2014) provide design recommendations to assess the "view out" of the window.

Nevertheless, there are fewer standards and guidelines available that consider the design of window view compared to other indoor environmental quality parameters such as, thermal comfort and air quality (ANSI/ASHRAE, 2017, 2019a). One reason for this might be that some standards are orientated more towards energy-efficiency goals rather than on occupant health and well-being, which is receiving growing scientific interest. Another reason may be the lack of a clear definition and method that can be used to assess “view quality”. Current standards provide partial recommendations but do not target every important aspect. For example, some building certification systems, such as LEED v4.1 (USGBC, 2020) and WELL v2-pilot (IWBS, 2019) provide design recommendations that evaluate the amount of occupied floor area that has a direct line of sight to the window – with optional criteria targeting other aspects of view quality. On the other hand, daylighting standards place more emphasis on the visual content seen by the occupants (CEN/TC 169, 2018; SLL, 2014), targeting specific features that may promote better view quality. Other guidelines focus more on urban planning, site layout, and the selection of an appropriate orientation of the building (Littlefair, 2011).

To help provide architectural design solutions that enhance occupant health and wellbeing, the multifaceted qualities of window views need to be identified. Scientific research is needed to build a framework that can be incorporated into building standards and green certification systems, which coalesces the fragmented literature found in many areas of research, including: architecture, urban planning, landscape, environmental psychology, and vision science. In fact, a recent study (Waczynska *et al.*, 2020) showed that current recommendations used to evaluate window views (CEN/TC 169, 2018) did not have a strong relationship with subjective appraisals given by observers, which suggests that design standards alone may not be sufficient when used to understand how people perceive window views. While this reinforces the need for a more comprehensive assessment method, other approaches also exist that either evaluate some, but not all important factors that influence overall view quality (Turan et al. 2019; Mardaljevic 2019; Li and Samuelson 2020), or can only be utilized post building construction (e.g., during post occupancy evaluations) rather than during the design phase (Hellinga and Hordijk, 2014; Matusiak and Klöckner, 2016). To help bridge these gaps, we have developed a comprehensive method to evaluate view quality that includes information derived from diverse sources.

We propose a framework that can be used to comprehensively evaluate the window view. Using a semi-systematic review style, we compiled current daylighting standards, green certification systems (Supplemental Material, Appendix A), and scientific literature into three underlying categorical variables, which were identified by recuring themes centred around the design process of façade and windows. We then introduce an index that demonstrates how the three variables are related to each other in the design process, and we also provide recommendations that could be used to define future research.

## 2. Window view assessment: Primary variables

In our framework, window view quality defines views that provide outdoor visual connection through the window, satisfying the physiological and psychological needs of building occupants. Underlying this definition are three primary variables that affect view quality: content, access, and clarity. We derived these three variables from our review of the design standards, while considering the design process of façade and windows (Fig. 1). Each design stage progressively requires more detailed levels of information: namely, 1) site selection, planning, and building massing; 2) facade and floorplan design; and 3) façade material selection and control.

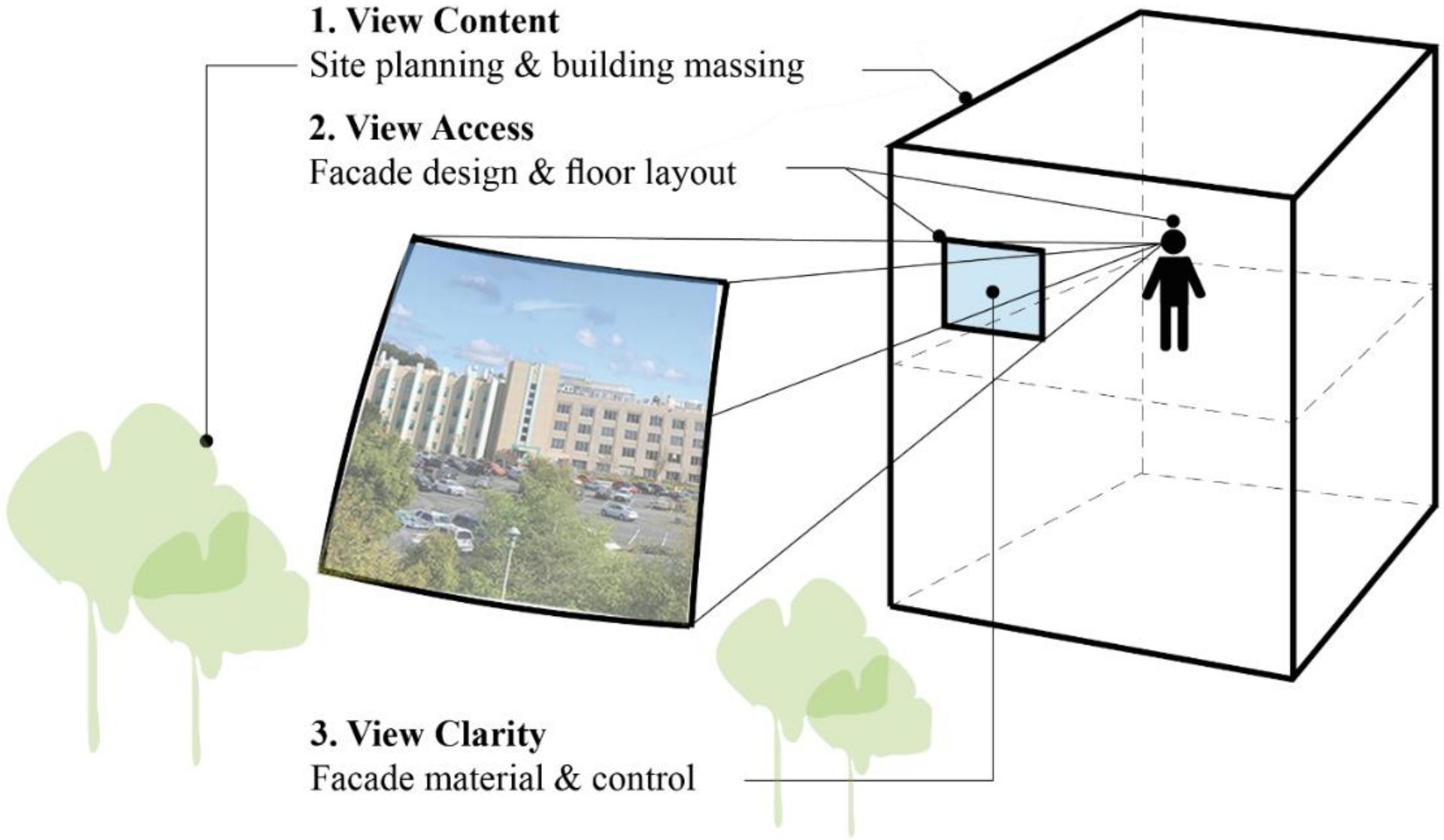

**Fig. 1**. The primary variables for driving view quality: content, access, and clarity.

We briefly define each of the three variables below. Although these variables appear as separate entities, we believe that they could also be related to each other.

- ***View Content*** is the sum of the visual features seen in the window view, for example, natural or urban features or the sky (Berman *et al.*, 2008; Grinde and Patil, 2009; Ulrich, 1981)).
- ***View Access*** is the amount of the view an occupant can see *from* the viewing position. Access primarily depends on the geometric relationships between the occupant and the window (Keighley, 1973a; Ne'Eman and Hopkinson, 1970).
- ***View Clarity*** addresses how clearly the content appears in the window view *when* seen by an occupant. It refers to both the design and the properties of the glazing and the shades that may change how the view is perceived through the window (Hill and Markus, 1968; Ko *et al.*, 2017; Konstantzos *et al.*, 2015).

Defining the three primary variables for view quality is the first of its kind. Table 1. Summary of existing studies of view assessment summarizes the aforementioned relevant literature published in the past eight years in comparison to the proposed framework. We believe that our framework can be applied in most buildings except for spaces which do not necessarily require views (e.g., museums and theaters), or where privacy is a problem (e.g., lavatories) (Ludlow 1976; Butler and Biner 1989). In other building spaces, including offices and classrooms, the windows' role of providing occupants with a view to the outside is more significant, and our framework was designed to evaluate these spaces.

Our framework is intended to guide a wide spectrum of building life cycle. For new constructions, it can guide their early design (e.g., visual content and initial window and floor layout) to construction stages (e.g., selection and installment of shading device). It can also be used to quantify the degree of view quality loss or enhancement caused by new construction (i.e., content shifts), floor or furniture layout changes (i.e., access shifts), and façade retrofitting, maintenance or operational change in shading devices (i.e., clarity shifts).

In our review, we first introduce each variable and relevant literature followed by building standards and green certification systems, which are then synthesized with empirical evidence derived from occupant assessments found in scientific research. Although literature came from different sources, they were each considered equally important within our review process. We only included literature available in English.

**Table 1.** Summary of existing studies of view assessment

| Reference | Basis | Primary view variables | | | Assessment |
|---|---|---|---|---|---|
| | | *Content* | *Access* | *Clarity* | |
| (Hellinga and Hordijk, 2014) | Review of scientific research and questionnaire study | • Nature (green or water)<br>• Horizontal layers<br>• Dynamic features (cars or traffic)<br>• Perceptual quality (diversity)<br>• Composition (dominant building(s))<br>• Building quality (maintenance, age, simple/complex design) | Viewing direction: perpendicular to window, Distance to window: 1.5 m | None | View quality score ranges 0 to 12<br><br>≥ 8: high<br>5 to 7: medium<br>≤ 4: low |
| (Matusiak and Klöckner, 2016) | Review of scientific research | • Nature (Greenery)<br>• Horizontal layers<br>• Content distance<br>• Weather (sky, cloudiness)<br>• Composition<br>• Aesthetical scene quality<br>• Individual difference (gender, age) | Horizontal view angle | Window design (fragmentation) | Quality of view<br><br>1: not satisfactory<br>2: satisfactory<br>3: good<br>4: very good |
| (Konstantzos and Tzempelikos, 2017) | Geometrical quantification | None | Solid angle (the portion of the occupant's visual field that is covered by a window) | View clarity index (VCI) | Effective Outside View (EOV) |
| (Purup *et al.*, 2017) | Review of scientific research | • Quality Factor (the relative subjective desirability of view areas in the reference view)<br>• Degree of privacy (risk of view from outside to the certain location in the room) | • Reference view: the unobstructed view from the top of the tower<br>• Limited view-out area projected onto the reference view are seen through windows<br>• Spatial assessment | None | View-Out Quality (0 to 1)<br>Degree of Privacy (0 to 1) |
| (Zanon *et al.*, 2019) | EN 17037 and LEED v4 | • Nature (flora, fauna, sky)<br>• Dynamic features (movement) | 75% of the floor of the rooms has a direct line of sight to the outdoors | None | "View Out" score (0-2 pts.) as a part of Qualitative Score (QS) for Visual Quality Index |
| | | • Content distance (≥ 7.5m) | | | |
| (Turan *et al.*, 2019) | EN 17037 and a principle of visual perception | • Nature (green space, water)<br>• Horizontal layers (sky, buildings and ground)<br>• Content distance (depth-of-field)<br>• Iconic landmarks<br>• Diversity | Spatial assessment | None | Total view potential: A cumulation of both the object count for each object type and the depth-of-field calculation for all of the rays (0 to 100 each). |
| (Mardaljevic, 2019) | BS 8206-2 and the aperture skylight index (ASI) | • Horizontal layers (sky, natural/man-made objects, ground) | Solid angle | None | View Lumen |
| (Li and Samuelson, 2020) | Geometrical and graphical quantification | Desirable/undesirable views defined by users<br>Optional weighting factors (e.g., vegetation, landmark) | Field of view (FOV) | None | View Score: a factor of both the sub-area of glass and its weighting factor as follows:<br>• -100: undesirable views<br>• 100: undesirable views<br>• 50: remained views |
| Proposed framework | Review of design standards, certification systems, and scientific research | • Nature and urban features<br>• Horizontal layers<br>• Content distance<br>• Dynamic features (movement)<br>• Perceptual and physical qualities | • View angle<br>• Spatial assessment<br>• Alternative view access | • Window design<br>• Glazing and shading materials<br>• Temporal attributes | View Quality Index (VQI), ranges 0 to 1<br>• VQI < 0.125: insufficient<br>• 0.125 ≤ VQI ≤ 0.375: sufficient<br>• 0.375 ≤ VQI ≤ 0.75: good<br>• VQI ≥ 0.75: excellent |

## 3. View content: What features?

In our framework, content is the sum of the visual features that are seen in the window view. Preferred views are often those containing visual features that connect occupants to the outdoor environment, while enhancing their aesthetical experiences and also support psychological restoration and general wellbeing (Matusiak and Klöckner, 2016).

Although most standards dictate the view can be evaluated from certain reference points (CEN/TC 169, 2018) or any point in the room (SLL, 2014), no further guidance is provided. Before the construction of the building, these approaches for evaluating view content may not be feasible. Instead, images of potential window views could be taken and their visual features could be analyzed to determine the quality of the content (Berto, 2005; Kent and Schiavon, 2020b; Ulrich, 1981; Ulrich *et al.*, 1991). Using images, Hellinga and Hordijk (Hellinga and Hordijk, 2014) proposed a view assessment method that asked respondents to evaluate certain visual features. Depending on their responses, weighted points to each question were tallied and a final view score was assigned to the image (e.g. low, medium or high quality view).

The European standard EN 17037 on daylighting (CEN/TC 169, 2018) and the Society of Light and Lighting, Lighting Guide (SLL-LG) 10 (SLL, 2014) show that window views can be assessed in terms of the content is that is seen by occupants (Table 2). The EN 17037 evaluates the quality of the view content with three levels of recommendations (minimum, medium, and high), whereas the SLL-LG 10 utilizes adjective labels to denote four levels (insufficient, sufficient, good, and excellent).

**Table 2.** Three levels of recommendations in the EN 17037 (EN) and the SLL-LG 10 for designing a window view (CEN/TC 169, 2018; SLL, 2014).

| **Level of view quality** | | **Number of layers** | | **Outside distance of the view (m)** | | **Horizontal sight angle (°)** | |
|---|---|---|---|---|---|---|---|
| EN | SLL | EN* | SLL | EN | SLL | EN** | SLL |
| - | Insufficient | - | Only sky or only foreground | - | < 6 | - | < 14 |
| Minimum | Sufficient | At least landscape layer | Landscape layer plus one other | ≥ 6 | | ≥ 14 | |
| Medium | Good | Landscape layer plus one other | | ≥ 20 | | ≥ 28 | |
| High | Excellent | All layers | All layers | ≥ 50 | | ≥ 54 | |

* Number of layers to be seen from at least 75% of utilized area

** For a space with room depth more than 4 m, it is recommended that the respective sum of the view opening(s) dimensions is at least 1.0 m x 1.25 m (width x height)

| **Environmental information criteria – EN (SLL)** | (Insufficient) | Minimum (Sufficient) | Medium (Good) | High (Excellent) |
|---|---|---|---|---|
| • **Location** | - | ✓ | ✓ | ✓ |
| • **Time** | ✓ | ✓ | ✓ | ✓ |
| • **Weather** | ✓ | ✓ | ✓ | ✓ |
| • **Nature** | - | - | ✓ (***) | ✓ |
| • **People** | - | - | (***) | ✓ |

*** Only one out of "nature" or "people" is required for a view quality of "Good" to be achieved.

**Table 3.** Window visual elements required in LEED v4.1 (USGBC, 2020) and WELL v2-pilot (IWBS, 2020).

| Certification | Visual elements |
|---|---|
| LEED v4.1 | • Views that include at least *two* of the following:<br>1) Flora, fauna, or sky;<br>2) Movement; and<br>3) Objects at least 7.5 m from the exterior of the glazing |
| WELL v2-pilot | • Views with a vertical view angle of at least 30° from occupant facing forward or sideways provide a direct line of sight to:<br>1) The ground or;<br>2) Sky |

While Table 2 shows that certain visual features are used to determine the level of view quality, LEED v4.1 (USGBC, 2020) and WELL v2-pilot (IWBS, 2019) provide optional criteria for visual elements that target view quality (Table 3). Below we have reviewed the literature that evaluates these and discusses how they influence view quality.

### *3.1. Nature and urban features*

Views with nature usually include scenes with greenery (e.g. trees) and these can have positive impacts on psychological well-being (van den Berg *et al.*, 2003; Berman *et al.*, 2008; Hartig *et al.*, 1996; Kaplan, 1993; Velarde *et al.*, 2007) and are desired by building occupants (Matusiak and Klöckner, 2016).

Design standards promote the incorporation of natural features (CEN/TC 169, 2018; MOHRUD, 2014; SLL, 2014). Flora or green space is one of the natural features specified by LEED v4.1 (USGBC, 2020) and WELL v2 (IWBS, 2020). Besides greenery, views with bodies of water (e.g., the sea, rivers, or lakes) (Kfir *et al.*, 2002) are also preferred. WELL v2 (IWBS, 2020) also recommends views to blue space. A study (White *et al.*, 2010) showed that both urban and nature views received higher affect ratings and perceived psychological restoration when they contained water (Ulrich, 1981; Ulrich *et al.*, 1991). Having a view of nature can also significantly increase the value of a building property, though mostly in residential buildings (Baranzini and Schaerer, 2011; Benson *et al.*, 1998; Bourassa *et al.*, 2004).

Many studies have demonstrated that views with natural elements are preferred by occupants over predominantly urban scenes (Grinde and Patil, 2009; Herzog, 1989; Kaplan, 1993, 1987; Ulrich, 1981). While this might imply that urban content is generally not preferable, some features – such as landmarks – are desirable in the window view (Baranzini and Schaerer, 2011; Damigos and Anyfantis, 2011). The age, maintenance, and complexity of urban design are features that influence how the window view is perceived (Baranzini and Schaerer, 2011; Hellinga, 2013; Matusiak and Klöckner, 2016).

### *3.2. Horizontal stratification*

Horizontal stratification refers to distinct boundaries seen across the horizontal axis of a view, creating visible layers between the ground, landscape, and sky (Markus, 1967). Each layer provides information about the local environment which may promote view quality. For example, the sky provides cues that may reveal the time of day and weather, the landscape shows the surrounding land, and the ground includes information on the floor (e.g. roads, pavements, and people) (Bell, 1973; Butler and Biner, 1989; Collins, 1976).

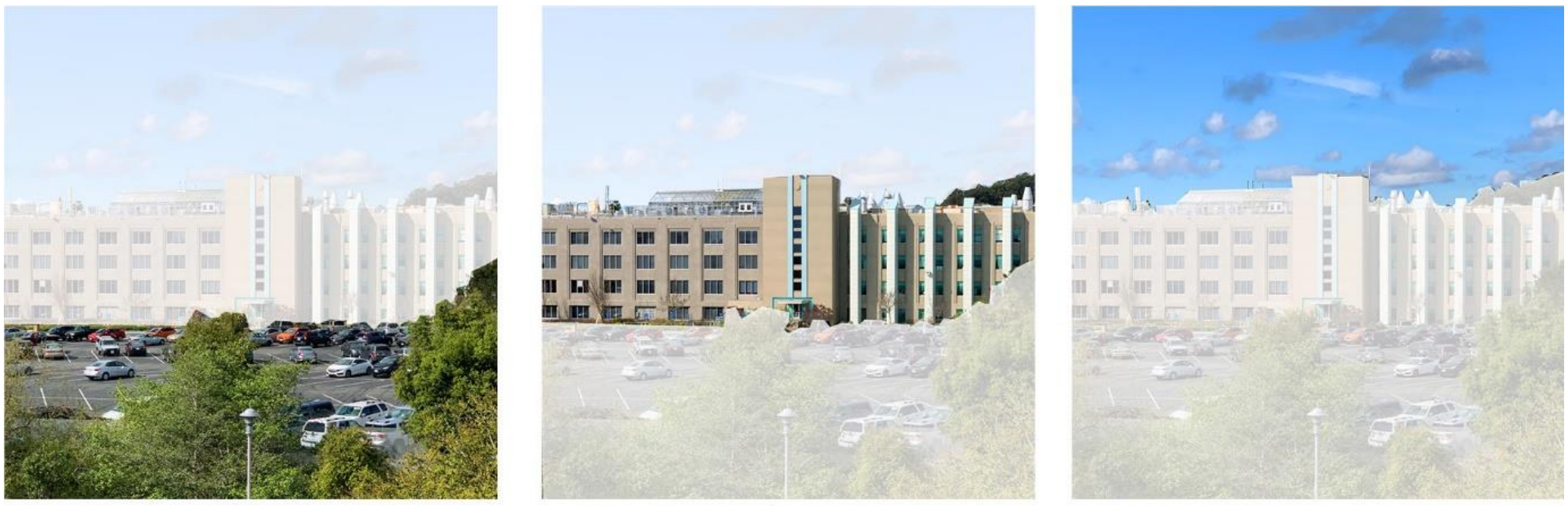

**Fig. 2.** A window view of a building in Albany, California that would be classed as "high" or "excellent" according to EN 17037 (CEN/TC 169, 2018) or the SLL-LG 10 (SLL, 2014) criteria in Table 2. Three horizontal layers have been identified: (a) ground layer, (b) landscape layer, and (c) sky layer.

The European standard (Table 2) evaluates views based on horizontal stratification (Fig. 2) and these have also been partially utilized in WELL v2-pilot (IWBS, 2019), which promote views containing the ground and sky layers (Table 3).

### *3.3. Content distance*

The distance between the window and outdoor object(s) also contributes to view quality (Kent and Schiavon, 2020a). Ne'Eman and Hopkinson (Ne'Eman and Hopkinson, 1970) suggested that when view content is far from the observer, it will relieve eye-strain (i.e. asthenopia) symptoms. In offices, this is a common problem when occupants are focused on their computer screens for prolonged periods, or when close objects are within the occupants' view field (Hayes *et al.*, 2007; Seguí *et al.*, 2015). Content seen from far away distances can help relax eye muscles after from viewing nearby tasks in an office or school (Henning *et al.*, 1997). Some green certification systems state this as the purpose of achieving their view credit (Supplemental Material, Appendix B).

The EN 17037 (CEN/TC 169, 2018) and SLL-LG 10 (SLL, 2014) use 6 m as the minimum acceptable distance for the outdoor object from a window, whereas LEED v4.1 (USGBC, 2020) utilize 7.5 m.

Although studies generally show that views with distant content are preferred (Herzog and Shier, 2016; Keighley, 1973b, 1973a; Markus, 1967), a recent study (Kent and Schiavon, 2020b) showed that this only applies when the content is predominantly urban. When the visual features include preferred elements described in Section 3.1, the effect of distance has a much smaller influence (e.g., occupants prefer nature to be seen closer in the window view). Another consideration is floor level, whereby views from higher stories are generally preferred (Olszewska-Guizzo *et al.*, 2018). However, this might be because distant content can be seen more easily.

### *3.4. Dynamic features (movement)*

Dynamic visual features (e.g. people and traffic) have been known to capture human attention (Orquin and Mueller Loose, 2013) and when seen in a window view, are generally preferred qualities (Butler and Biner, 1989; Hellinga, 2013). This also include changes that occur naturally across the time of day or year, which can also influence visual preferences (e.g., daylight patterns, weather, and season variations such as tree foliage) (Brooks *et al.*, 2017; Duffy and Verges, 2010). Even views to roadwork and building construction were preferred, as long as the noise or vibration are not intrusive (Musselwhite, 2018). Dynamic visual information may occur in different layers of the window view (Section 3.2). The EN 17370 (CEN/TC 169, 2018) and SLL-LG 10 (SLL, 2014) specify these under environmental information criteria, whereas LEED v4.1 (USGBC, 2020) also specifies these under view quality.

Although we have yet to find any study that has systematically evaluated "dynamic features" as a design parameter, it is likely to be influenced by the quality and scale of the moving object as well as the duration, and frequency of the dynamic features in the window view (Milosavljevic *et al.*, 2012; Zoest *et al.*, 2004). Although our approach is somewhat subjective, dynamic features are those that cause noticeable changes in the view (e.g., vehicles, clouds), whereas other content that has less obvious movement (e.g., small birds, waving flags) could be excluded for brevity. The distance of the movement to the observer may also contribute to a view quality. For example, nearby movement (e.g., pedestrians passing in front of the window) might be distracting but distant movement (e.g., traffic flow from a further distance) might be more preferable (Ludlow, 1976).

## 4. View access: How much?

View access is a metric quantifying how much of the window view(s) an occupant can see through from a particular location within a space. In most commercial buildings, open-plan layouts are commonly used (Oldham and Rotchford, 1983; Yildirim *et al.*, 2007), and creates a challenge in providing suitable view access for all the occupants. Access encompasses problems of understanding how the distance between the view and occupant influences minimum window size requirements (Aries *et al.*, 2010; Ne'Eman and Hopkinson, 1970) and identifying the availability of view in a shared space.

There are variations to how standards account for view access. We predominantly focus on how much can be seen based on the geometric relationships between the occupant and their window view(s), for example, the distance to and size of the openings. However, standards and literature both provide different methods for evaluating access.

### *4.1. View angle of a window*

The angles from the occupant's viewing position to the vertical or horizontal perimeters (e.g., the frame bars) of a window are known as the viewing angles. The EN 17037 (CEN/TC 169, 2018) and SLL-LG 10 (SLL, 2014) use the view angle to evaluate the size of a window (Table 2). They specify that the view should have a horizontal view angle equal to or higher than $14^{o}$ from the occupants' viewing position to meet the "minimum" design recommendation. The minimum recommended view angles in the literature vary considerably (Table 4), which not only vary by the horizontal angle size (i.e., the width of view window(s)) (CEN/TC 169, 2018; SLL, 2014), but also, the vertical view angle (i.e., the height of view window(s)) (IWBS, 2019), or the smaller of the two view angles (USGBC 2019).

Preferred view angles can also be influenced by the visual features in the view. The minimum acceptable horizontal angle of windows tends to be smaller in distant views (Ne'Eman and Hopkinson, 1970), since large features appear smaller from a far distance. Keighley (Keighley, 1973b) showed that the preferred window size was influenced by the horizontal stratification of the content, whereby the position of the skyline boundary in the view influenced the preferred height of the window and its frame.

**Table 4.** Thresholds for the view angles of window view(s) from a given position.

| Minimum view angle | View content | References |
|---|---|---|
| Horizontal view angle of 14° | Landscape view with sky or ground | (CEN/TC 169, 2018; SLL, 2014) |
| Vertical view angle of 30° | Sky or ground view | (IWBS, 2019) |
| The smaller view angle of 11° | Landscape view (no nature) | (Heschong Mahone Group, 2003; USGBC, 2019, p. 4) |
| The smaller view angle of 9° | Landscape view with nature | (Heschong Mahone Group, 2003; USGBC, 2019, p. 4) |

Depending on which view angle is used, the approaches in Table 4 provide different results. For example, an occupant who is seated away from a shallow strip window would have a large horizontal view angle, but the vertical view angle would indicate poor access to the view. Therefore, there is no general agreement over approaches used in Table 4.

#### *4.1.1. View Factor*

View Factor is an assessment method that considers both the content and visual angle of the view, which was originally developed by the Heschong Mahone Group (Heschong Mahone Group, 2003) and was later adopted in LEED v4.1 (USGBC, 2020) and WELL v2-pilot (IWBS, 2019). Table 5 identifies the corresponding view angles and a description of the view, whereby one denotes the lowest and five the highest levels of the view factor. From the occupant position, the view angle is first calculated by taking the smaller of either the vertical or horizontal angle, and then, the visual elements are evaluated when the view factor is finalized. If the physical dimension of the window(s) or furniture layouts are not available, it may be difficult to determine the view factor. In addition, the use of the smaller angle between the vertical and lateral for the threshold view angle can oversimplify the actual available area of some windows.

**Table 5.** View Factor calculation based on the view angle and content (Heschong Mahone Group, 2003).

| View Factor | View angle (º) | Description | Compliant |
|---|---|---|---|
| 5 | 50 to 90 | Completely filled the visual field of the observer seated at the workstation | Yes |
| 4 or 5 | 40 to 50 | 4 (non-nature view) or 5 (nature view) | |
| 4 | 20 to 40 | Filled about one-half of the visual field | |
| 3 or 4 | 15 to 20 | 3 (non-nature view) or 4 (nature view) | |
| 3 | 11 to 15 | Filled about one-half of view factor 4, but still with a coherent view | |
| 2 or 3 | 9 to 11 | 2 (non-nature view) or 3 (nature view) | Yes/No |
| 2 | 11 to 15 | A narrow and typically fractured view | No |
| 1 or 2 | 4 to 11 | 1 (non-nature view) or 2 (nature view) | |
| 1 | 1 to 4 | A glimpse of sky or sliver of the outside environment | |

### *4.2. Distance from a window and window-to-wall ratio*

In BREEAM (BRE Global Limited, 2018), view access is promoted by showing that relevant building areas are within 8 m of an external wall containing a window. In LEED v4.1 (USGBC, 2020), one of the view access criteria is to provide unobstructed views located within the distance of three times the head height of the window. Considering that the typical ceiling height of an office building is 2.7 m, the distance is also approximately 8 m. Similarly, WELL v2 (IWBS, 2020) and DIN 5034 (DIN, 2011) requires that workspaces are within 10 m of a window.

The window size and resultant view access have also been determined by the distance between the workstation and the window. The BREEAM (BRE Global Limited, 2018) utilizes the window-to-wall ratio (WWR), which is the ratio of the window area to the gross area of the surrounding wall. Depending on the distance in open-plan offices, these values vary from: 20 % (< 8 m), 25 % (8 - 11 m), 30 % (11 - 14 m), and 35 % (> 14 m).

Beside distance from the window, WWRs are also found in building energy performance standards (ANSI/ASHRAE, 2019b) and widely appear in the literature (Hardy and O'Sullivan, 1967; Marino *et al.*, 2017; Ochoa *et al.*, 2012; Tzempelikos and Athienitis, 2007). However, since various factors influence window size satisfaction, the preferred WWR can vary considerably (e.g. 50 % to 80 % (Ludlow, 1976) or 44% to 100% (Dogrusoy and Tureyen, 2007)). This suggests that other parameters also influence the preferred window size. Roessler (Roessler, 1980) showed that the preferred width and size of a window depend on visual features that could be seen. Other studies have shown that the amount of visible sky, which is dependent on the distance the occupant is from the window, can significantly influence the view satisfaction (Abd-Alhamid *et al.*, 2020; Keighley, 1973b).

### *4.3. Alternative view access*

When the view angle of a window is too small or the distances from the window is too far, designers may also find it challenging to provide occupants with access to a view. In these cases, green certification systems provide alternative methods of visual access for the occupied space (Table 6). Internal courtyards and atria that contain natural elements (Section 3.1) are common design solutions (BRE Global Limited, 2018; DGBC, 2010; NGBC, 2016). Green Mark (BCA Green Mark, 2010) and Green Star (NZGBC, 2016) standards promote the integration of nature onto surfaces such as green walls and roof gardens (Fig. 3).

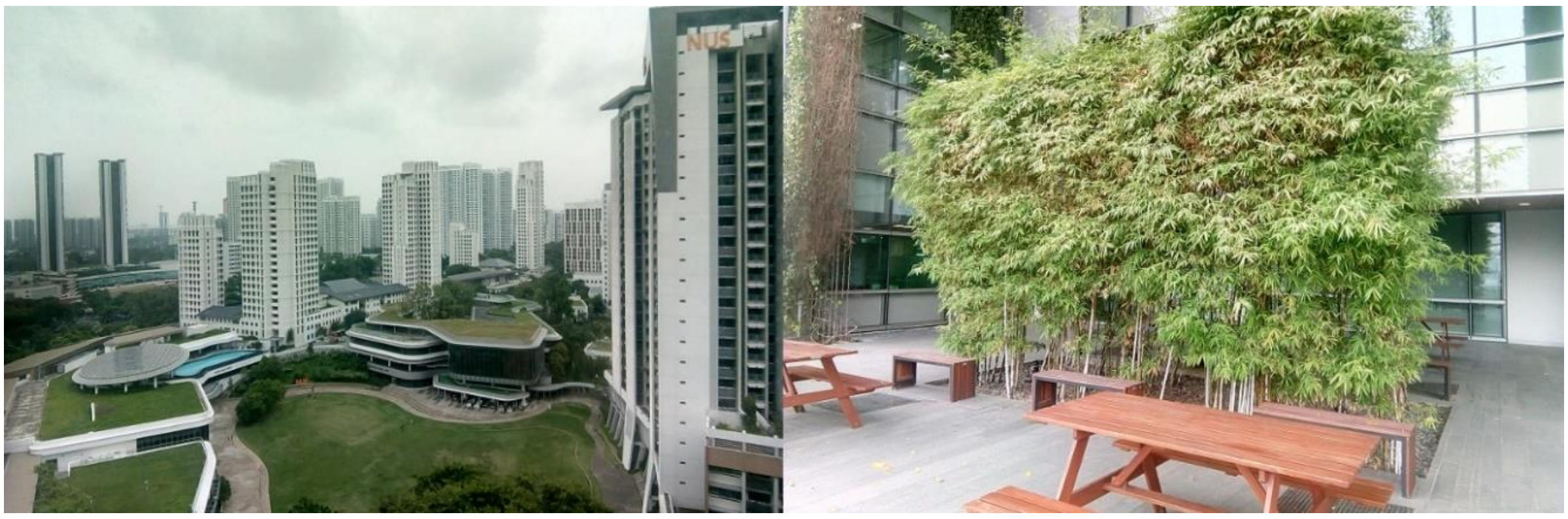

**Fig. 3.** View of the National University of Singapore (NUS) campus that contain a green-roof design (left) and an outdoor sky garden showing integrated nature in the CREATE Tower Building (right).

**Table 6.** Alternative methods of providing visual access out of the occupied space in green building certifications. This describes the requirements of the visual content that need to be in a view from the courtyard or atrium.

| Certification | View access | Criteria | Visual features |
|---|---|---|---|
| BREEAM NOR **(NGBC, 2016)**; BREEAM NL **(DGBC, 2010)**; BREEAM **(BRE Global Limited, 2018)** | Courtyard or atrium | Minimum distance of visual content of at least 10 m. | Greenery, plant containers, etc. |
| Green Globes **(ECD Energy & Environment Canada Ltd., 2004)** | Atrium | From all primary interior spaces. | Provides a view to the building exterior |
| Green Star **(NZGBC, 2014)** | | Atria must be at least 8 x 8 m in width and depth at any viewing position inside room. | High quality internal view to the eye of the building occupant. |
| Green Star NZ **(NZGBC, 2016)** | | Atrium must be at least 8 m in width from any viewing position inside room. | Not specified. Promoting visual interest to distant objects to allow eye rest. |

### *4.4. Spatial assessment of view access*

In green certification systems, the requirements of view access mostly rely on a spatial assessment, which is the percentage of floor space that can provide visual contact with the window(s). Table 7 presents the list of green building certifications that provide credits based on the spatial assessment method. Within the regularly occupied spaces, predefined criteria must be met and these vary depending on the green certification system. For example, BREEAM (BRE Global Limited, 2018) specifies that 95 % of the total floor areas must provide an "adequate" view out, whereby adequacy is defined by a view that contains a landscape and not only the sky layer at a seated eye-level. The DGNB system used in Germany evaluates the visual contact with the outside by specifying that all office rooms need to have a direct link to the outside and at least 80 % of all break and social rooms (DGNB GmbH, 2018). A limitation to the spatial assessment approach is that most systems do not require a view angle that specifies how much is enough for the visual contact.

Table 7. Green building certifications and the credit distribution for designing spaces with the spatial assessment.

| Certification | Credits allocated to the design | | | Verification of credit(s) |
|---|---|---|---|---|
| | *1* | *2* | *3* | |
| BERDE **(PHILGBC, 2017)** | 50 | 75 | - | Floor plans and models showing spaces with direct outdoor views |
| BREEAM UK: 2018 **(BRE Global Limited, 2018)** | 95 | - | - | Entire building areas show that 95 % of total floor area provides an adequate view out. |
| BREEAM International **(BRE Global Limited, 2015)** | 80 | 95 | - | Design and as-built drawings, sectional drawings, site inspection report and photographs, and confirmation from contractor or design team. |
| Green Building Index **(GBI, 2009)** | 60 | 75 | - | Demonstrate direct lines of sight to the windows at height of 1.2 m from floor level. |
| GreenShip **(GBC Indonesia, 2012)** | 75 | - | - | Direct view of windows when straight line path is drawn from viewing position. |
| Green Star NZ **(NZGBC, 2016)** | 60 | 90 | - | Reference drawings with calculations, design and built phases architectural drawing. |
| HQE: 2014 **(Cerway, 2014)** | 30 | 50 | 75 | Architectural drawings with percentage calculations and site visit. |
| IGBC V3 **(IGBC, 2014)** | 75 or 95 | - | - | Achieve direct line of sight at 2.1 m from floor level to window. |
| LEED Canada: 2009 **(CAGBC, 2010)** | 90 | - | - | Plan view and sectional view drawings. Feasible furniture layout to interpret analysis. |
| LEED India: 2011 **(IGBC, 2011)** | 90 | - | - | Plan view and sectional view drawings. |
| LEED V4.1 **(USGBC, 2020)** | 75 | - | - | Unobstructed views. |
| Pearl Rating System for Estidama **(Emirate of Abu Dhabi, 2010)** | 75 | - | - | Intent show in project design and plan and section drawings. |
| WELL V2-pilot **(IWBS, 2019)** | 50 | - | - | Architectural drawings. |
| WELL V2 **(IWBS, 2020)** | 75 | - | - | Architectural drawings. |

LEED v4.1 (USGBC, 2020) also uses another method of evaluating access, which measures the floor area that has access to the view(s) from multiple lines of sight (i.e., laterally 90° apart from each other at the viewing position of the occupant). If this criterion is achieved, the design will provide different view content, which may provide a more diverse overall environment. It also increases the probability of providing a least one functional view (e.g., if one area is overlit, it may be possible to preserve at least one view to the outside when other windows are shaded by blinds). The method does not specify in detail how to identify the number of different types of views or the aggregate amount of view available. This information would help designers plan window and floor layouts, as it is common to have multiple windows in one or more viewing directions in offices (Collins, 1976).

## 5. View clarity: How clear?

View clarity is a metric assessing how clearly the visual content in the view can be seen by the occupant when considering the visual obstructions present at the window – before, after, or inside the glazing layer(s). The concept of visual clarity can generally be related to a "*sense of satisfaction*" (Aston and Belichambers, 1969) when using the term to indicate a "*cleaner, more vital*" or "*visually distinct and clear*" appearance (Boyce, 1970). These definitions were used to assess the impact of electrical lighting on visual performance and can be adapted to view clarity. In our framework, view clarity encompasses the glazing system, window design, the glazing and shading materials (both external and internal), visual acuity, contrast sensitivity and color perception. While certain aspects of view content and access change over time (e.g., seasonal changes in foliage, and furniture arrangement), clarity is expected to vary on a much regular basis. Dynamic changes in the outdoor environment, especially sunlight transmission through the window, influence shading requirements that protect occupants from glare and thermal discomfort.

Daylighting standards and green certification systems require that the glazing material of a view window should provide a clear, neutrally colored, and undistorted by frits, fibers, patterned glazing, or added tints (CEN/TC 169, 2018; USGBC, 2020). Similarly, ASHRAE 189.1 requires undiffused glazing with a haze value (ASTM International, 2013) less than 3 % for view fenestration (ANSI/ASHRAE/ICC/USGBC/IES, 2017). Such recommendations can conflict with the design considerations for the glazing support (e.g., mullions) and occupant comfort (e.g. shading devices), which may obstruct the visual connection to the outside. When the structural elements or shades are present in the window, observers may be dissatisfied with the fractured visual information (Wilson, 2005) or the unnatural changes to the color (Clear *et al.*, 2006). This reduces the quality of the view. Careful placement of window mullions and a balance between daylight entering and the view seen outside the building is therefore important (Kim and Kim, 2010).

### *5.1. Window design*

The design of the glazing supports such as the window frame and mullions can have an important role in how the view is perceived. The shape of a window frame matters (Dogrusoy and Tureyen, 2007) as it may block or shift the boundaries of outdoor elements that provide occupants with visual cues in the view. The preferred horizontal and vertical proportions of window frames vary depending on the content of the view (Keighley, 1973b; Markus, 1967; Ne'Eman and Hopkinson, 1970).

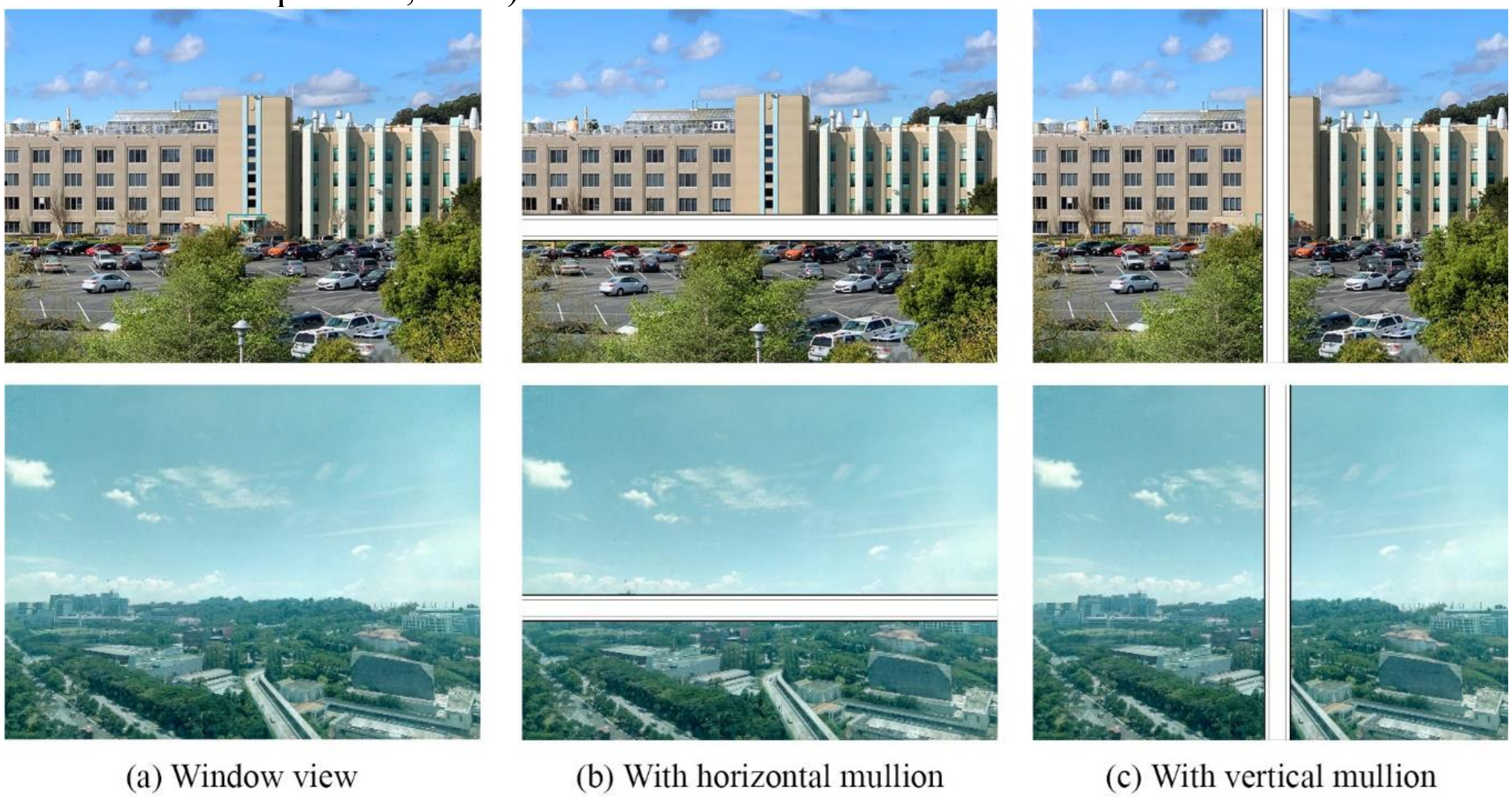

**Fig. 4.** Window views with (a) ground, landscape, and sky layers and without mullions, the same window view but with hypothetical (b) horizontal and (c) vertical mullions.

The visual features in a view can be partially or completely distorted when blocked (Markus, 1967) as demonstrated in Fig. 4, whereby we present three images of window views in Albany, California (top) and at the NUS (bottom). This shows three scenarios: unobstructed view (left), horizontal (center) and vertical (right) mullions. When placing a mullion along the horizontal axis, this may distort the clarity of the visual content at certain viewing positions (e.g. seated at workstation). The distortion is prominent in the NUS view since the landscape and skyline boundary is blocked. When the mullion is placed along the vertical axis, the clarity of the visual content is somewhat retained. While this occurs because the horizontal layers contrast against the vertical mullion, stereoscopic vision also allows humans to see the content either side of the mullion (i.e., the visual parallax effect), which does not occur when the mullion is horizontal.

Both view content and access can influence clarity. Keighley (Keighley, 1973a) showed that the view satisfaction decreased as the number and width of mullions increased. In a distant view, wider mullions created a large decrease in the satisfaction due to more obstructed features in the view (Fig. 4-b). The phenomenon of reducing the clarity of a window view through mullions or other obstructions was also in the study (IJsselsteijn *et al.*, 2008; Wilson, 2005), whereby window elements fracture the content and this may change how the occupant perceived the visual information.

### *5.2. Glazing and shading materials*

With the advances in envelope materials and shading systems (e.g., fabric shades, electrochromic glass, etc.), various façade designs are increasingly common. Although these are designed to reduce the risks of glare, overheating (Carmody *et al.*, 2004; Sun *et al.*, 2018) or privacy issues (Dogrusoy and Tureyen, 2007; Sundstrom *et al.*, 1982), they also block or distort the window view (Collins, 1976; SLL, 2014) as shown in Fig. 5-a. Depending on the optical properties of the shading system, the clarity of the view content seen in the window can be drastically reduced (Fig. 5-b).

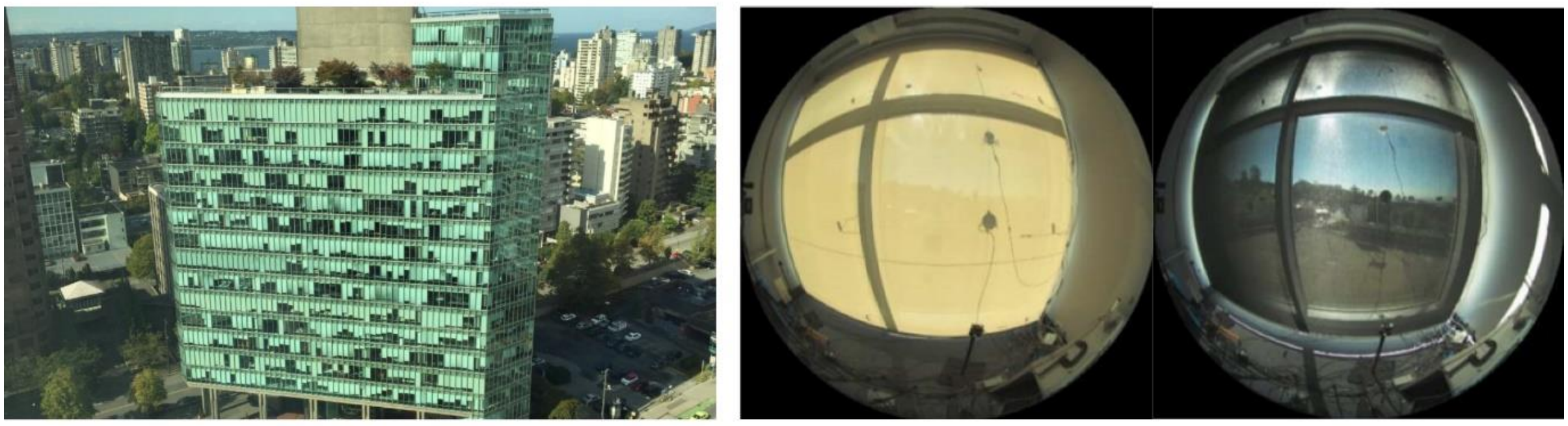

(a) A building in Vancouver

(b) Window view with two different fabric shades, LBNL Window Testbeds (71T)

**Fig. 5.** (a) Floor-to-ceiling glass wall with the high level of the window shade occlusion (Mulpuri, 2020), (b) different view clarities of a window view with two fabric shades (Ko *et al.*, 2017).

In the case of tinted glass, such as electrochromic glass, the color of an outdoor view can be shifted and it may reduce the contrast sensitivity and color perception of the observer (Ko *et al.*, 2017). For example, blue tinted windows may influence how the occupant perceive the content of a window view. Compared to clear and neutral class, the sky layer on clear days could look darker and it may shift its visibility of the landscape and ground layers depending on the color composition of each. Although the effect of tinted glass on the occupants was studied in a few studies, these are mainly focused on the "colored light" (i.e., daylight with the Correlated Color Temperature distortion coming through the tinted glass) on the occupants' visual task performance, subjective indicators (e.g., sleepiness) (Liang *et al.*, 2021) or thermal perception (Chinazzo *et al.*, 2018). The effect of the color-naturalness of a window view can be studied further.

#### *5.2.1. View Clarity Index*

The openness, directional transmission (i.e., transmission related to the geometric optics, without diffusion or redirection, CEN/TC 33, 2008), and shape (e.g., the weaving pattern and angle or orientation of the openings) of building envelope materials can influence view clarity (Hill and Markus, 1968). Some studies assessed the impact of the optical and physical properties of the façade materials on the clarity of a window view (Ko *et al.*, 2017; Konstantzos *et al.*, 2015) and a study (Konstantzos *et al.*, 2015) has led to the development of the view clarity index (VCI) of a given fabric shade, calculated using Eq. (1):

$$VCI = 1.43\,(OF)^{0.48} + 0.64\left(\frac{OF}{T_v}\right)^{1.1} - 0.22 \qquad (1)$$

whereby, OF refers to the openness factor (%) and $T_v$ is the visible transmittance of the fabric shade (%). This index was derived by investigating how well observers could see through 14 different shading fabrics in a test room. The VCI predicts the clarity of a fabric shade using an empirical model based on the result. VCI values range from zero (perfectly diffused shading) to one (no shading).

### *5.3. Temporal and compositional attributes of view clarity*

Although many characteristics vary and influence view clarity (e.g. daylight, glare, privacy, etc.), it is important to consider the temporal and compositional characteristics of view clarity: how long a window can be viewed under which levels of clarity. Shading devices, by its nature, changes the view clarity of windows over time depending on its operation. Understanding the duration, frequency (i.e., temporal attributes) of view clarity changes and the level of obstruction (i.e., compositional attributes) caused by the shading systems are two primary factors for view clarity. Although it is not clear what minimum acceptable levels could be used, it may be possible to inform these thresholds from climate-based daylighting metrics, such as the Spatial Daylight Autonomy (sDA) and Annual Sunlight Exposure (ASE) (Daylight Metrics Committee, 2012). Both metrics assess the behavior of daylight during standard operating hours (Reinhart *et al.*, 2006), and ASE specifies the percentage of time the floor area exceeds 1000 lux for 250 hours, thereby measuring the potential of glare and overheating that triggers the use of blinds. The occlusion of the shading systems will resultantly influence view clarity. The use pattern of shading device is a critical factor to predict the view clarity but it is difficult to make generalization about typical occlusion values given the multiple aspects (i.e., orientation, sky condition, season, time of day) affecting the occlusion. In the 50 buildings data, the south façade occlusion was most commonly 40 to 70% and 15 to 25% for the north facades (Van Den Wymelenberg, 2012). Typical adjustment frequency, 'rate of change' is another aspect to consider for the use pattern of shading device. The frequency is heavily influenced by the types of control mechanism, especially for internal shading systems, such as manual versus automated controls. Shading devices manually controlled by the occupants have a lower daily rate of change (e.g., 0 % to 15%) compared to automated shading devices (Reinhart and Voss, 2003). Internal shades are often partially lowered (Fig. 5-a) to maintain some level of visual connection to the outside, while simultaneously protecting the occupants from glare and/or thermal discomfort (Konis, 2013).

Therefore, the dynamics of shading systems should be considered while developing overall building acceptability regarding view quality. The occlusion of the shading systems changes the luminance contrast between outdoor and indoor environments and the view clarity. This can help designers to plan the façade, its materials, and controls.

## 6. View Quality Framework

### *6.1. Summary*

Fig. **6** summarizes the view quality framework based on the three variables (content, access, and clarity). For each variable, we highlight a set of criteria and proposed design considerations based on the literature. The general importance placed on each criterion was based upon consensus found across design standards and scientific research and to further consolidate our work, we discussed the resultant framework with several design practitioners as part of a rigorous internal review process (Supplemental Material, Appendix C).

| Variable | Criteria | Design considerations |
|---|---|---|
| **View content (*Vcontent*)** | | |
| 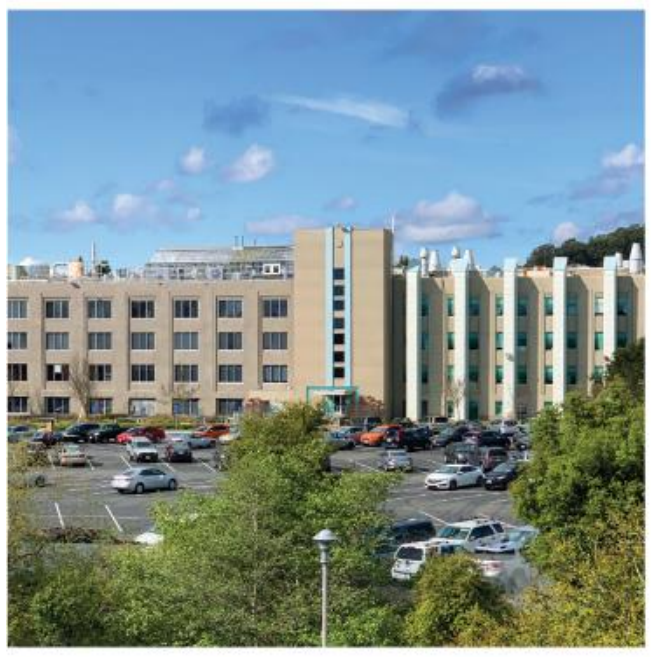 | Nature and urban features | • Position windows to face greenery and/or and natural water features when possible |
| | Horizontal | • Ensure at least landscape and/or sky layer |
| | Content distance | • Avoid nearby content (< 6 m from window) in view particularly in urban content |
| | Dynamic features (movement) | • Provide views showing dynamic features (≥ 6 m from window) (e.g., traffic flow and people) |
| **View access (*Vaccess*)** | | |
| 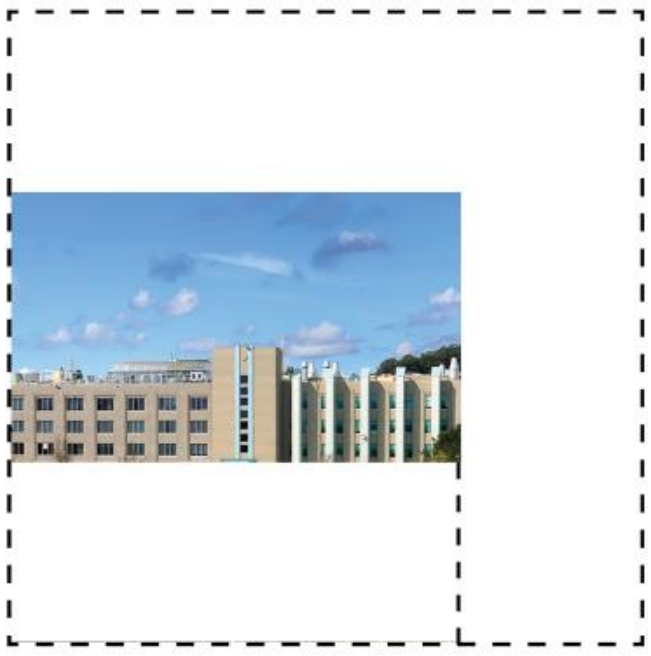 | View angle | • Achieve the minimum view angle*<br>• Achieve View Factor of at least 3 |
| | Alternative access | • Design atrium and courtyard (at least 8m in width) with pleasant visual elements (e.g., nature) |
| | Spatial assessment | • At least 75 % of the floor area has direct access to window view ( ≥ minimum view angle*) |
| **View clarity (*Vclarity*)** | | |
| 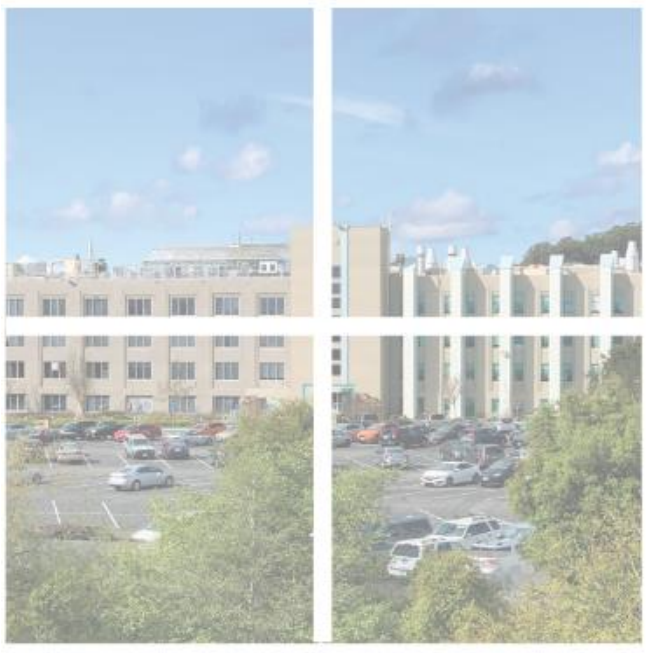 | Window design | • Avoid placing the horizontal/vertical mullions at the boundaries of the ground, landscape and sky layers |
| | Glazing and shading materials | • Select shading material considering the VCI* when available |
| | Temporal attributes | • Control the glazing and shading systems to provide the desired clarity of window view(s) (≥ minimum number of occupied hours* that exceed minimum view clarity* for the space) |

* Design parameters that we could not find consensus in the reviewed standards, green certificate systems, and scientific literature. Section 7 discusses the future research.

**Fig. 6.** View quality framework listing the three variables and criteria we have identified, and the design considerations that may need to be implemented.

### *6.2. Introduction to the View Quality Index*

While providing an organizing framework for unifying existing standards and future research is important, we should also consider how the framework and research findings can be used in the design process. Therefore, we propose a conceptual index, the view quality index (VQI), by incorporating previous works on view assessment from the literature. The VQI expresses the relationship among the primary variables that others may use to build upon as some aspects require further research. We believe that researchers may benefit from considering the VQI when developing research questions and methods on window view.

All three variables, content, access, and clarity are important for view quality and interrelated. For example, if the view content is of low quality (i.e., a nearby concrete wall), the occupant would not be satisfied even with high levels of view access and clarity. By the same token, if an occupant sits far from the window (i.e., insufficient view access), the view would not be satisfactory regardless of how good the view content and view clarity are. Similarly, a view that constantly requires a shading device to protect occupants from glare achieves little view clarity and therefore insufficient view quality. Fig. **7** illustrates three examples of unsatisfactory view conditions due to the low quality of one of the variables even if the other two are high.

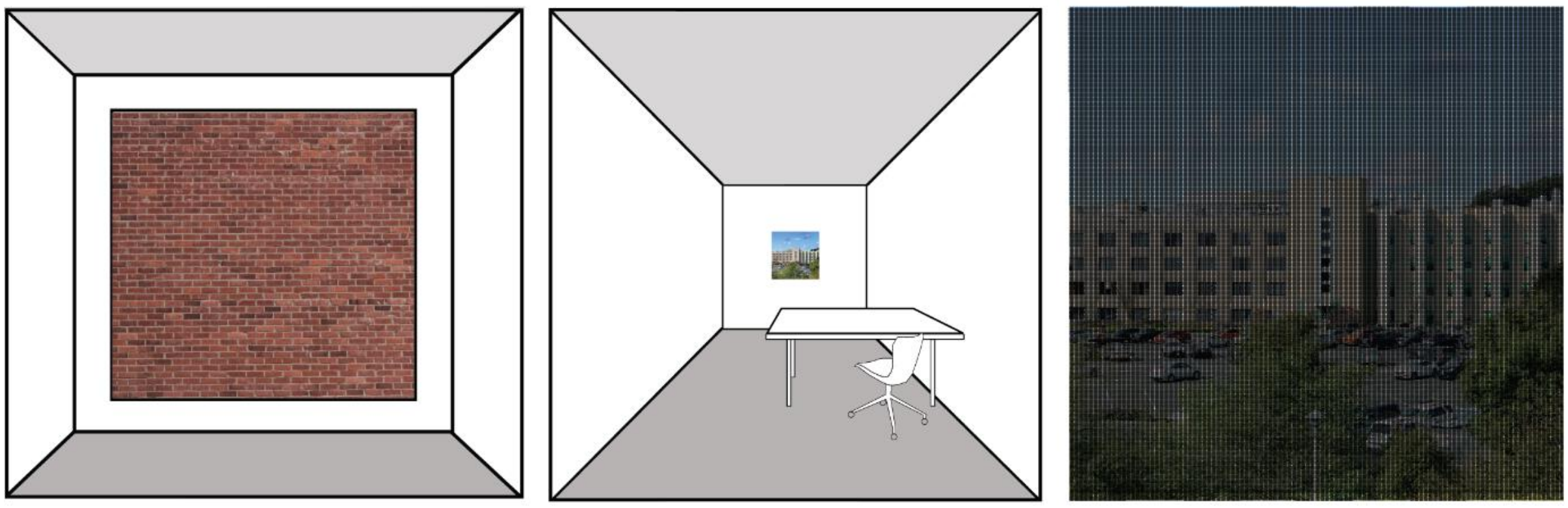

**Fig. 7.** The relationships between view content, view access, and view clarity.

Based on this concept, we define the relationship between the primary variables to quantify view quality of a given window using Eq. (2):

$$VQI = V_{content} \cdot V_{access} \cdot V_{clarity} \quad (2)$$

whereby, we assume each variable is equally weighted. Although unequal weightings are possible in different contexts, designers can easily emphasize one parameter (see Supplemental Material, Appendix E), which could be informed through practical knowledge or other design requirements. For example, in ground floor office spaces, more priority may need to be given to clarity to ensure privacy requirements are met when occupants perform clerical tasks, meaning that access and content – while are still important, require less emphasis. When occupant movement is limited (e.g., patient rooms in hospital or prison cells), limited opportunities may be available to help maximize access and clarity (due to stringent environmental requirements), therefore designers may prioritize view content.

When there are multiple windows, openings can be treated as one large window with a continuous view if the distance between them is small. Similar recommendations can also be found in section C.3. of the EN 17037 (CEN/TC 169, 2018), but the separation distance is not specified nor is it apparent whether the view will still be perceived as a single entity when multiple windows are within a close proximity of each other. Research has shown that when the view angle is greater than 60°, parts of the window view are outside from the central area of the occupant's visual field (Ne'Eman and Hopkinson, 1970). Since we cannot substantiate this to multiple windows, our general approach is designed for individual windows. In order to deal with distantly placed windows that occupy the same building space (e.g., corner offices with multiple windows separated by long perpendicular walls), more research – as specified in Supplemental Material, Appendix G(research needs) – is required.

We assign 0 (less than the minimum acceptable range), 0.5 (equal to the minimum acceptable range), and 1 (greater than or equal to the saturation range) for each variable. The determination of each parameter and their ranges are as follows.

$V_{content}$ takes on a value that represents the available number of view layers, the outside content distance (i.e., the median distance from the nearest to farthest identified objects measured from the window), presence of dynamic feature(s) and nature features in the window view (Section 3), which is derived from Eq. (3):

$$V_{content}=L_{sky}+L_{landscape}\cdot wf_{ct.dis.}+L_{ground}\cdot wf_{movement}+L_{nature}\cdot wf_{nature}$$

with

$$L_{sky}, L_{landscape}, L_{ground}, L_{nature} = \begin{cases} 0.25 \text{ if present in the scene} \\ 0 \text{ if absent in the scene} \end{cases},$$

$$wf_{ct.\ dis.} = \begin{cases} 1 \text{ if } 50\ m < \text{Content distance} \\ 0.75 \text{ if } 20\ m < \text{Content distance} \leq 50\ m \\ 0.5 \text{ if } 6\ m < \text{Content distance} \leq 20\ m \\ 0 \text{ if Content distance} \leq 6\ m \end{cases}$$ except natural features (see Section 3.1),

$$wf_{movement} = \begin{cases} 1 \text{ if distant dynamic feature}(s)(>6\ m) \text{ is present} \\ 0.5 \text{ if no dynamic feature}(s)(\leq 6\ m) \text{ is present} \\ 0 \text{ if nearby dynamic feature}(s)(\leq 6\ m) \text{ is present} \end{cases}$$

$$wf_{nature} = \begin{cases} 1 \text{ if } \%\text{ of natural features in the scene} > 50\ \% \\ 0.75 \text{ if } 25\ \% < \%\text{ of natural features in the scene} \leq 50\ \% \\ 0.5 \text{ if } \%\text{ of natural features in the scene} \leq 25\ \% \\ 0 \text{ if no natural feature in the scene} \end{cases} \quad (3)$$

whereby, $L_x$ represents the horizontal layers (sky, landscape, and ground) and nature present in the view. We assign content distance and the presence of dynamic features as weighting factor ($wf_x$) to the landscape layer (with the exception of extraordinarily preferred features such as, landmarks, see Section 3.1) and ground layer in the scene as we believe that the two visual features closely related to the corresponding layers (e.g., people and vehicles would mostly be in the ground layer). We also assign the weighting factor ($wf_{nature}$) to the nature layer. The thresholds for the percentage of natural features occupying the scene are used to calculate $V_{content}$ in the examples in Supplemental Material, Appendix F, but they should be further validated by an additional study. $V_{content}$ range from zero (i.e., the lowest possible rating such as a scene nearby a wall) to one (i.e., the highest rating such as sky, distant landscape, ground layers with nature and dynamic features).

$V_{access}$ represents the view angle scores that can be derived from the $V_{content}$ (Section 3) and can be calculated using Eq. (4):

$$V_{access} = \begin{cases} 1 \text{ if } \alpha_{view} \geq \alpha_{saturation} \\ y \text{ if } \alpha_{min} < \alpha_{view} < \alpha_{saturation} \text{ with } y = \frac{1}{2}\left(\frac{\alpha_{view}}{\alpha_{saturation} - \alpha_{min}}\right) \\ 0.5 \text{ if } \alpha_{view} = \alpha_{min} \\ 0 \text{ if } \alpha_{view} < \alpha_{min} \end{cases} \quad (4)$$

whereby, $\alpha_{view}$ represents the viewing angle of the window as seen from an observer's position on the floorplan (assuming visual gaze is on the window view), $\alpha_{min}$ represents the minimum view angle needed to achieve an acceptable level of view access, and $\alpha_{saturation}$ is the largest view angle wherein the observer does not need any further increase in the outdoor view. $y$ is the $\alpha_{view}$ falling between $\alpha_{min}$ and $\alpha_{saturation}$, that is rescaled by setting $\alpha_{min}$ as 0.5 and $\alpha_{saturation}$ as 1. $V_{access}$ ranges from zero (i.e., smaller than the minimum angle threshold) to one (i.e., larger than the saturation angle). Supplemental Material, Appendix D summarizes the view angle thresholds in the literature. These values are based on different experimental settings and across a limited set of conditions. Further research is necessary to derive more generalized thresholds.

$V_{clarity}$ represents the view clarity scores based on the VCI (Section 5) and can be calculated according to Eq. (5):

$$V_{clarity} = \begin{cases} 1 \text{ if } \beta_{clarity} \geq \beta_{saturation} \\ y \text{ if } \beta_{min} < \beta_{clarity} < \beta_{saturation} \text{ with } y = \frac{1}{2} \left( \frac{\beta_{clarity}}{\beta_{saturation} - \beta_{min}} \right) \\ 0.5 \text{ if } \beta_{clarity} = \beta_{min} \\ 0 \text{ if } \beta_{clarity} < \beta_{min} \end{cases} \quad (5)$$

whereby, $\beta_{clarity}$ represents the clarity of a window view as seen from an observer's position on the floorplan, $\beta_{min}$ represents minimum view clarity needed to achieve an acceptable level of view clarity, and $\beta_{saturation}$ represents maximum view clarity, wherein the observer does not need any further increase in the clarity of the outdoor view. $y$ is the $\beta_{clarity}$ falling between $\beta_{min}$ and $\beta_{saturation}$, that is rescaled by setting $\beta_{min}$ as 0.5 and $\beta_{saturation}$ as 1. $V_{clarity}$ ranges from zero (i.e., completely blocked window) to one (i.e., clear and unobstructed window). As discussed in Section 5, further research is needed to identify the minimum and saturation thresholds that achieve sufficient visual contact.

Fig. 8 (using a case study) and Supplemental Material, Appendix F (using the empirical findings from a study) show how (1) and (2) can be used to calculate VQI and $V_{content}$ across different window views, whereby the range of values have been labeled according to the criteria used in the SLL-LG 10 (SLL, 2014): Insufficient (< 0.125), Sufficient (0.125 ≤ View Quality ≤ 0.375), Good (0.375 ≤ View Quality ≤ 0.75) and Excellent (≥ 0.75), respectively.

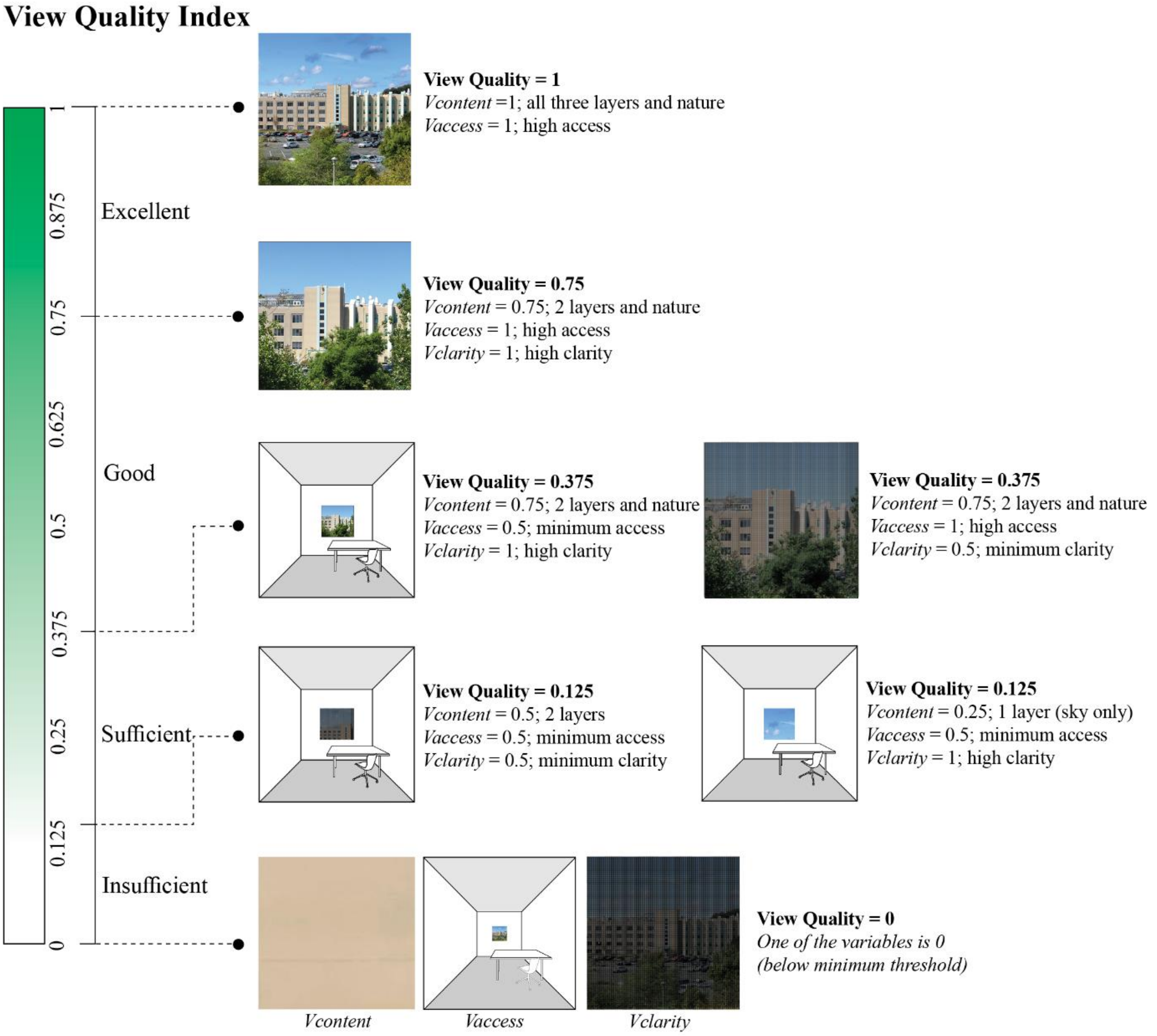

**Fig. 8.** Conceptual calculations of view quality.

## 7. Limitations and research gaps

Our literature review aims to coalesce a range of literature into a window view quality framework and to help inform design practice. However, some limitations and gaps in our understanding need to be acknowledged.

### *7.1. View content*

Since view content is heavily subjective, quantifying an evaluation metric is inherently difficult. It is unclear how some visual features in the window view are classified. For example, if trees are seen nearby or far away, they will produce different levels of visible nature in the view, but the actual visual features seen would remain the same. When mixing other features (e.g., urban) in the view, it is not clear if the view content in both cases should be classified as nature or mixed. Because only a few methods are available to serve this purpose, content assessment often relies on a degree of subjective interpretation. By a similar token, the relationship between the number and type of visual features and overall visual appraisal has also yet to be explored. According to the gestalt theory of perception, the whole is greater than the sum of its parts (Ellis, 2013). On this basis, the accumulation of certain attributes (e.g., greenery or horizontal layers) seen in the window may be not linearly related to the overall quality of the window.

In cases when the visual features are diverse and mixed, these approaches are not necessarily useful. Whereas literature (Markus, 1967) and standards (CEN/TC 169, 2018; SLL, 2014) both promote views with three-layers (i.e. ground, landscape, and sky), it is not clear how much of each should be present in the view. Although we think the composition between the layers should be relatively balanced, threshold preventing one layer from dominating the other two (e.g., 90 % landscape, 5 % for both sky and ground) might be useful.

Another way to improve the content assessment could be to analyze a view through the use of perceptual or physical measures. These are different from the aforementioned approaches used to characterize view content (Section 3). Instead of identifying individual features in the view, the general content (i.e., sum of all features) is measured in one overall assessment. While granularity in the assessment of content is somewhat sacrificed, this could provide a simpler and more accurate measure. However, such approaches have not yet been clearly developed (Velarde *et al.*, 2007).

Perceptual qualities, such as the preference matrix proposed by Kaplan and Kaplan (Kaplan and Kaplan, 1989), is an approach used to understand how visual features are perceived according to its coherence, legibility, complexity, and mystery. In Kaplans' theory, sufficient levels of coherence and legibility are necessary to understand the information in the visual scene, but these must be balanced by enough complexity and mystery to yield a visual preference (Herzog, 1989; Herzog and Leverich, 2003). Though its practicability should be carefully examined (Stamps, 2004), it ventures out a new dimension of view research. Other researchers also found that some perceptual attributes of views such as openness (Collins, 1976; Herzog and Chernick, 2000; Ozdemir, 2010) may influence view quality.

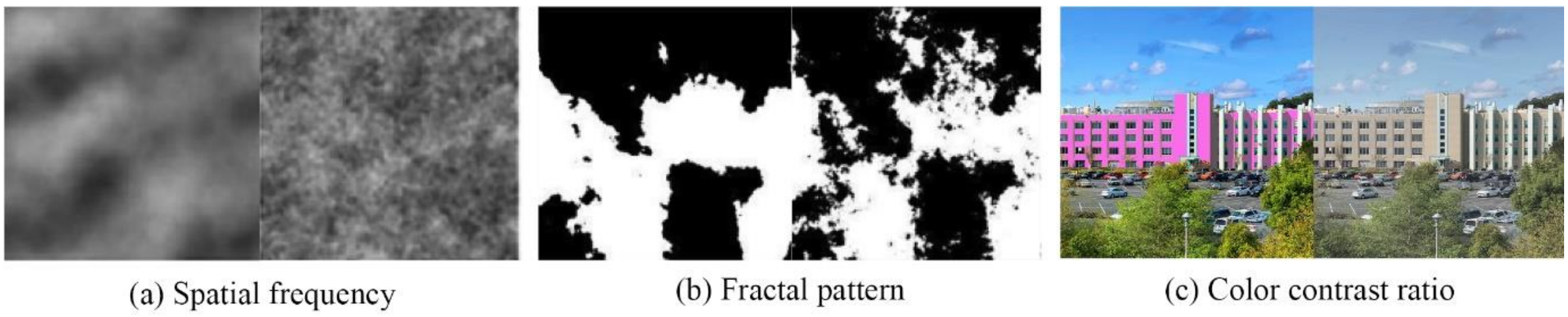

**Fig. 9.** Baseline images (left) versus preferred images (right) in terms of (a) spatial frequency (Juricevic *et al.*, 2010), (b) fractal patterns (Abboushi *et al.*, 2019), (c) color contrast ratio.

The physical qualities of the visual scene could be the basis for measuring its visual features. Scenes containing nature have a specific spatial frequency (Fig. 9-a), which produce visual patterns of luminance that allows the human eye to process efficiently and comfortably (Wilkins, 2016). This physiological basis may also relate to the human preference for nature scenes. Similar to spatial frequency, fractal patterns commonly found in nature may have a relevant impact on visual responses. Fractal images (Fig. 9-b) consist of self-similar patterns with varying magnification scales. Fractal images facilitate higher visual preference (Abboushi *et al.*, 2019) and promote restorative effects compared to non-fractal images (Hagerhall *et al.*, 2008). Color contrast ratio (Fig. 9-c) and chromaticity from sunlight also influence the subjective and physiological responses to a view (An *et al.*, 2016; Beute and de Kort, 2013; Wilkins, 2016). The larger the separation of two colors (i.e., the larger the color difference) in the CIE 1976 diagram (ISO, 2019), the greater the visual discomfort (Haigh *et* *al.*, 2013). With the advancement in the field of computer vision (Forsyth and Ponce, 2002), designers could analyze the physical qualities of potential view images taken from the project site, which can later be used in the building design process.

### *7.2. View access*

Although there is some consensus that defines view content, view access is not as comparably well-defined. For example, view access has been evaluated using view angles, viewing distances, and spatial assessment metrics. While these existing measures partially capture relevant information, we believe a more reliable indicator is necessary that unifies the segmented requirements in current recommendations may help designers to evaluate their designs with view accessibility.

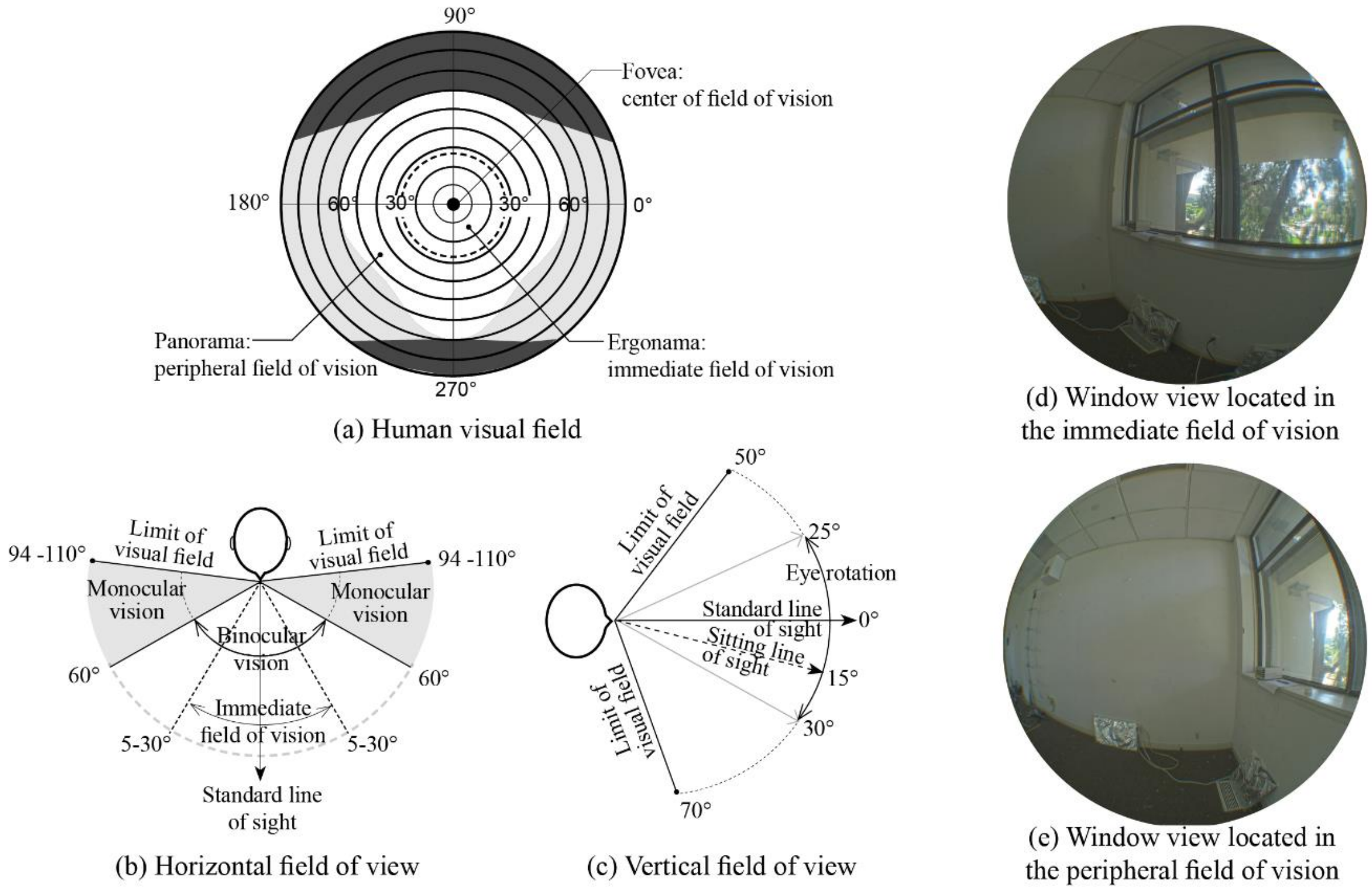

**Fig. 10.** (a) Human field of vision and view angles that can also be seen for the (b) horizontal and (c) vertical visual fields adapted from (Boring *et al.*, 1948; SLL, 2014), window views located in the climatic chamber at UC Berkeley showing (d) the immediate field of view and (e) the peripheral field of view.

Using both distance from and size of a window, solid angles or a percentage (i.e., how much of the view occupies the observer’s visual field) could be calculated. While some further studies need to verify the usefulness of solids angles, it could also be an essential factor in considering view access index. Besides solid angles, the viewing direction of the occupant (i.e., whether looking directly at the view, or concentrating their gaze onto a task) and the location of the view in their visual field (Fig. 10) can be considered. One important question is whether an observer still appraises the window view in the same manner even when they are not focusing on it. A study by Stone and Irvine (Stone and Irvine, 1994) evaluated view access in an experiment by facing observers towards (direct) and perpendicular (indirect) to the window. The results showed that observers sometimes benefited more from the visual stimulation from the window when they directly looked at the view. In early design phase (e.g., before the furniture layout is set), the designers may consider quantifying the possibilities of a quality view within a 360° visual field (i.e., all possible viewing directions).

The spatial metrics that most green certificate systems use for access assessment require more detailed predefined criteria, such as specifying minimum view angles and a method of evaluating access from multiple windows. When multiple views are present in an occupant's visual scene, further studies are required to determine whether these should be aggregated or treated separately in the assessment of view access. A study (Magistrale, 2014) found that higher satisfaction ratings were given to three windows when rating the "amount" of windows, although the area of openings – from one or two windows – was the same.

### *7.3. View clarity*

Although there is some information available that can be used to evaluate clarity (e.g., VCI), more research is needed to understand how clear window views appear under a wider range of scenarios (e.g., different shading and tinted glazing materials). In a study that investigated the visual performance of electrochromic windows and fabric shading systems, view clarity was reduced when the optical properties of the electrochromic glass lowered the transmission of sunlight (Ko *et al.*, 2017). Assessing visual clarity under a wide range of test conditions (e.g., indoor and outdoor (day)lit conditions) and implementing a variety of visual performance tests (e.g., color-distance, contrast sensitivity, object recognition) (Chichilnisky and Wandell, 1995; Pelli *et al.*, 1988) could increase the accuracy and applicability of the current model (Konstantzos *et al.*, 2015). Since some of the conditions are also dynamic, there is a need to establish minimum design thresholds for visual clarity and duration (i.e., the percentage of the occupied hours when view clarity is at an acceptable level) for both fully or partial occluded scenarios.

Once the view clarity assessment is developed further, it can be simultaneously analyzed with daylight availability and visual discomfort for the optimized design and control of glazing and shading devices. In a recent study, Garreton et al. (Yamín Garretón *et al.*, 2021) used VCI, Useful Daylight Illuminance, and Daylight Glare Probability to assess the relationship between outdoor view, daylight availability, and glare protection given by nine shading fabrics. Although their results cannot be applied to a wider range of window blinds or systems, it emphasizes the need for more research that can refine assessments given to view clarity, allowing for the design of windows that better meet the visual needs of occupant.

### *7.4. VQI: weightings, the spatial and temporal assessments*

One aspect which may require further investigation is the weightings placed on each variable. We assumed that three variables are equally weighted (i.e., content, assess, clarity, and their individual criteria are balanced). A similar method was used previously (Hellinga and Hordijk, 2014), but the parameters (e.g., nature or horizontal stratification) are not weighted equally. This raises questions about which parameters are more important.

Further steps to expand upon our approach could be to include spatial (Section 4.4) and temporal (Section 5.3) qualities. The quality of a window view varies depending on the occupant's position within the floorplan (Yildirim *et al.*, 2007), time of day and year (Brooks *et al.*, 2017), and the type of task or activity that the occupant is performing (Stone and Irvine, 1994). There are some differences in the degree to which variable is influenced by these factors, but we believe that all three are impacted and should, therefore, be addressed.

Once the spatial and temporal assessment thresholds for overall building acceptability are defined, they can be simultaneously analyzed with other environmental quality factors for the integrated design (Ko *et al.*, 2018; Purup *et al.*, 2017).

## 8. Conclusion

We propose a framework and a view quality index derived from the findings and recommendations found in building standards, green certification systems, and scientific literature that provides designers and researchers with a comprehensive approach for evaluating window view quality. We introduced three primary variables: content, access, and clarity. For each variable, we provide the current state-of-the-art, their limitations as well as existing research gaps and further research directions. We also described how to quantify them and how they are related.

Carefully designed window views can enhance the physiological and psychological well-being of the building occupants. With increasing emphasis placed on promoting the positive human impact of the built environment, the demand for sophisticated design metrics and tools for window views will increase. To meet this demand, we need to provide a more comprehensive assessment framework that meets the holistic requirements of occupants and propose new directions of study that can bridge the gaps between current research and design.

## Acknowledgments

This study was supported by the Center for the Built Environment (CBE), at the University of California, Berkeley; and the Republic of Singapore’s National Research Foundation through a grant to the Berkeley Education Alliance for Research in Singapore (BEARS) for the Singapore-Berkeley Building Efficiency and Sustainability in the Tropics (SinBerBEST) Program. We would like to thank Lisa Heschong, Peter Tregenza, and the CBE industry partners for their feedback on the manuscript.

## Funding details

The authors report no funding.

## Disclosure statement

The authors report no declarations of interest.

# Supplement Materials
## Appendix A

List of the green building certifications that integrate view in their guidelines according to their version, date of release, country, and the credit that it belongs to.

| Certification (Version) | Date | Country | Credit |
|---|---|---|---|
| BEAM Plus (V1.2) | 2012 | Hong Kong | SA 9 Neighbourhood Daylight Access |
| BEAM Plus (V2.0) | 2018 | Hong Kong | SS 4 Neighbourhood Daylight Access |
| BERDE GBRS (V2.0.0) | 2017 | Philippines | EQ-04 Visual Comfort |
| BREEAM | 2018 | United Kingdom | HEA 01 Visual comfort |
| BREEAM International | 2015 | International | HEA 01 Visual comfort |
| BREEAM-NL | 2010 | Netherlands | HEA 2 View out |
| BREEAM-NOR | 2017 | Norway | CN 12 and CN 13 |
| BREEAM-SE | 2013 | Sweden | HEA 2 View out |
| CityLab (V2.0) | 2018 | Sweden | 10 Lighting |
| DGNB System | 2019 | Germany | 3 Visual connect with the outside |
| Green Building Evaluation Label (China Three Star) | 2014 | China | 8.2.5 View outside |
| Green Building Challenge 2002: GBTool User Manual | 2002 | International | S4.1: Visual access to the exterior from primary occupancies |
| Green Building Index (V1.0) | 2009 | Malaysia | EQ12 External views |
| Green Globes | 2004 | Canada | Daylighting |
| Green Mark (GM RB) | 2016 | Singapore | (i) Biophilic design |
| GreenShip (V1.1) | 2012 | Indonesia | IHC 4 Outside view |
| GreenShip (V1.2) | 2013 | Indonesia | IHC 4 Outside view |
| Green Star AU | 2017 | Australia | - |
| Green Star Kenya | 2014 | Kenya | IEQ 8 External views |
| Green Star NZ: Office Design & Built | 2009 | New Zealand | IEQ 10 |
| Green Star NZ (V3.1) | 2016 | New Zealand | IEQ 10 |
| Green Star SA | 2014 | South Africa | Views and lines of sight |
| G-SEED | 2016 | South Korea | 7.1 |
| HQE | 2014 | France | 10.1.2. Having access to outside views in sensitive spaces |
| IGBC Green New Building Rating System (V3.0) | 2014 | India | IEQ Credit 3 |
| LEED (V4.1) | 2020 | United States | Quality views |
| LEED Canada | 2009 | Canada | - |
| LEED India | 2011 | India | IEQ Credit 8.2 Daylight and views – Views |
| Minergie | 2019 | Switzerland | - |
| Pearl Building Rating System for Estidama | 2010 | Emirate of Abu Dhabi | - |
| WELL (V2-pilot) | 2019 | United States | L05 Enhanced Daylight Access: Part 3 Ensure Views |
| WELL (V2) | 2020 | United States | M09 Enhanced Access to Nature: Part 1 Provide Nature Access Indoors |

## Appendix B

List of green building certifications that state that the purpose of awarding credit(s) to the design of window views is to reduce visual eye strain within the indoor environment.

| **Green certification** | **Purpose description of credit(s) awarded to design of view** |
|---|---|
| BREEAM-NL | To encourage adequate provision of an external view in all relevant workplaces. This is in order to prevent eye strain and break the monotony of the indoor environment. |
| BREEAM-NOR | Adequate view out to reduce eye strain and provide a link to the outside. |
| BREEAM-SW | To allow occupants to refocus their eyes from close work and enjoy an external view, thus reducing the risk of eyestrain and breaking the monotony of the indoor environment. |
| Green Building Index V1.0 and V3.0 | To reduce eyestrain for building occupants by allowing long distance views and provision of visual connection to the outdoor environment, which include greenery and/or water bodies. |
| GreenShip V 1.1 and V1.2 | To reduces eye fatigue by providing long-distance views provides a visual connection to the outside of the building. |
| Green Star NZ and Green Star NZ V3.1 | To encourage and recognise reduced eyestrain for building occupants by allowing long distance views and the provision of visual connection to the outdoors. |

## Appendix C

Since our work was heavily informed by existing and mainstream literature, we presented our framework to approximately 150 architects and building designers during an invitation only conference managed by our research group. The purpose of this exercise was to gather feedback from the perspective of a practitioner, who are the target audience that would translate our framework into the building design. While this was the general aim of these discussions, it was not to derive new results for our framework.

## Appendix D

Thresholds for the view angles of the window views in a given position.

| $V_{content}$ | **View content** | $\alpha_{minimum}$ | $\alpha_{saturation}$ | **References** |
|---|---|---|---|---|
| 0.25 | Sky or ground view | Vertical view angle of 30° | - | (IWBS 2019) |
| | Landscape view (no nature) | The smaller view angle of 11° | The smaller view angle of 90° | (Heschong Mahone Group 2003) |
| 0.5 or higher | Landscape view with nature | The smaller view angle of 9° | The smaller view angle of 50° | (Heschong Mahone Group 2003) |
| | Landscape view with sky or ground | Horizontal view angle of 14° | Horizontal view angle of 54° | (CEN/TC 169 2018) |

## Appendix E

Example VQI calculation when unequal weightings are assumed. In Section 6.2, (2) assumed each variable in the VQI formula is equally weighted. If unequal weightings preferred, designers can use (6) by allocating different coefficients to each variable:

$$VQI = k_{content}V_{content} \cdot k_{access}V_{access} \cdot k_{clarity}V_{clarity} \quad (6)$$

whereby, $k_x$ represents the coefficient of each variable (content, access, and clarity). The product value of the three $k_x$ should equal to one. For example, if two is assigned to $k_{content}$, the product of $k_{access}$ and $k_{clarity}$ should be 0.5 (i.e., each should be the square-root of 0.5, which equals 0.71).

## Appendix F

Example of Vcontent calculations for eight window views (M.G. Kent and Schiavon 2020).

| Window view image | $V_{content}$ | Values |
|---|---|---|
| 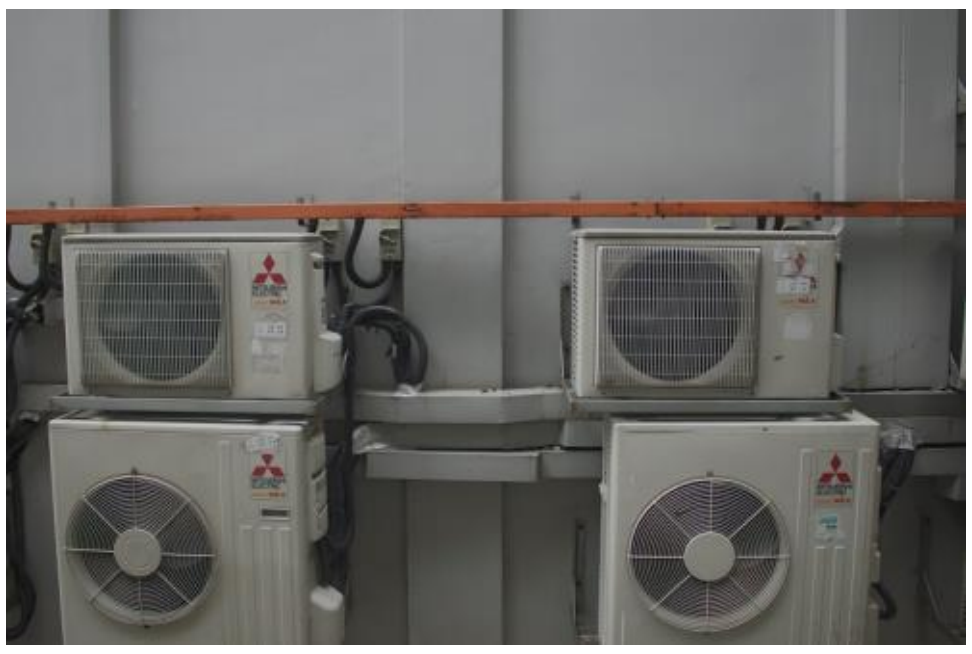 | $L_{sky}$ | 0 (no sky layer) |
| | $L_{landscape}$ | 0.25 |
| | $wf_{ct.dis.}$ | 0 (OLD 2) |
| | $L_{ground}$ | 0 (no ground layer) |
| | $wf_{movement}$ | 0.5 (no regular movement) |
| | $L_{nature}$ | 0 (no nature) |
| | $wf_{nature}$ | NA |
| | $V_{content}$ = 0 + 0.25 · 0 + 0 · 0.5 + 0 = 0 (**Insufficient**) | |
| 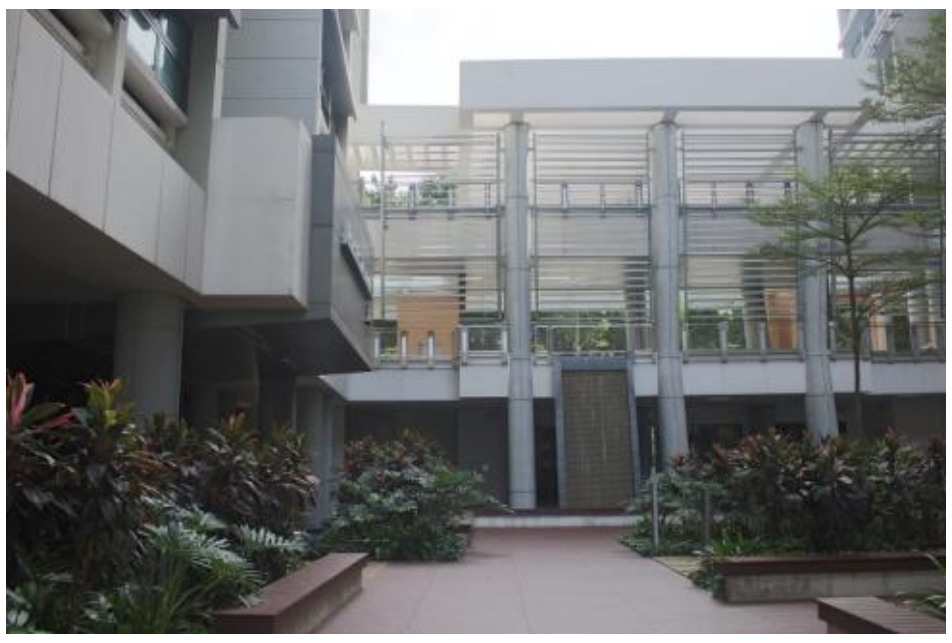 | $L_{sky}$ | 0.25 |
| | $L_{landscape}$ | 0.25 |
| | $wf_{ct.dis.}$ | 0.5 (OLD 14) |
| | $L_{ground}$ | 0.25 |
| | $wf_{movement}$ | 0 (nearby movement) |
| | $L_{nature}$ | 0.25 |
| | $wf_{nature}$ | 0.5 (< 25 %) |
| | $V_{content}$ = 0.25 + 0.25 · 0.5 + 0.25 · 0 + 0.25 · 0.5 = 0.5 (**Good**) | |
| 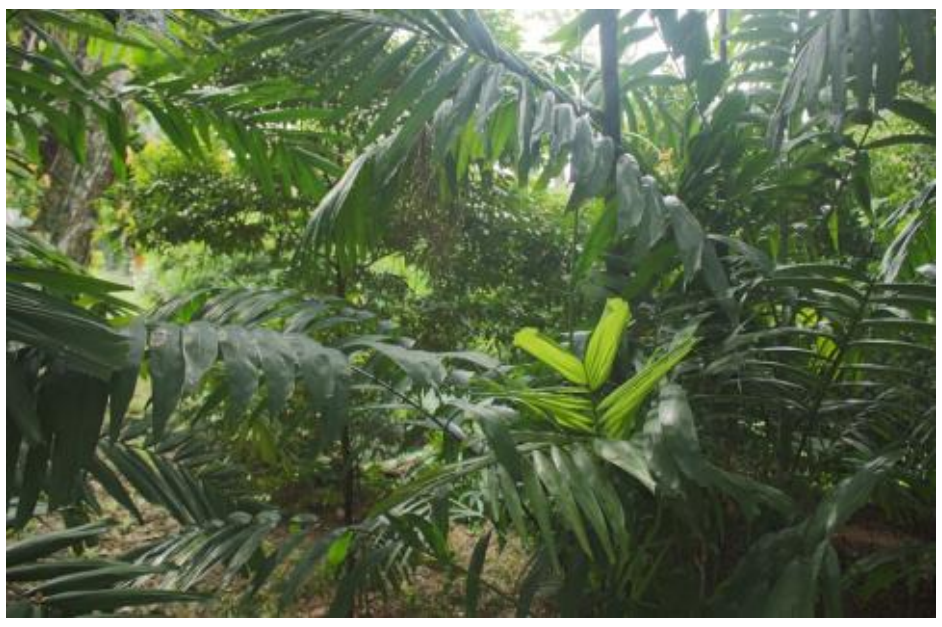 | $L_{sky}$ | 0 (no sky layer) |
| | $L_{landscape}$ | 0.25 |
| | $wf_{ct.dis.}$ | 1 (OLD 2, nature) |
| | $L_{ground}$ | 0 (no ground layer) |
| | $wf_{movement}$ | 0.5 (no regular movement) |
| | $L_{nature}$ | 0.25 |
| | $wf_{nature}$ | 1 (> 50 %) |
| | $V_{content}$ = 0 + 0.25 · 1 + 0 · 0.5 + 0.25 · 1 = 0.5 (**Good**) | |
| | $L_{sky}$ | 0 |
| | $L_{landscape}$ | 0.25 |
| | $wf_{ct.dis.}$ | 0.75 (OLD 31) |
| | $L_{ground}$ | 0.25 |
| | $wf_{movement}$ | 1 (regular movement) |
| | $L_{nature}$ | 0.25 |
| | $wf_{nature}$ | 1 (> 50 %) |

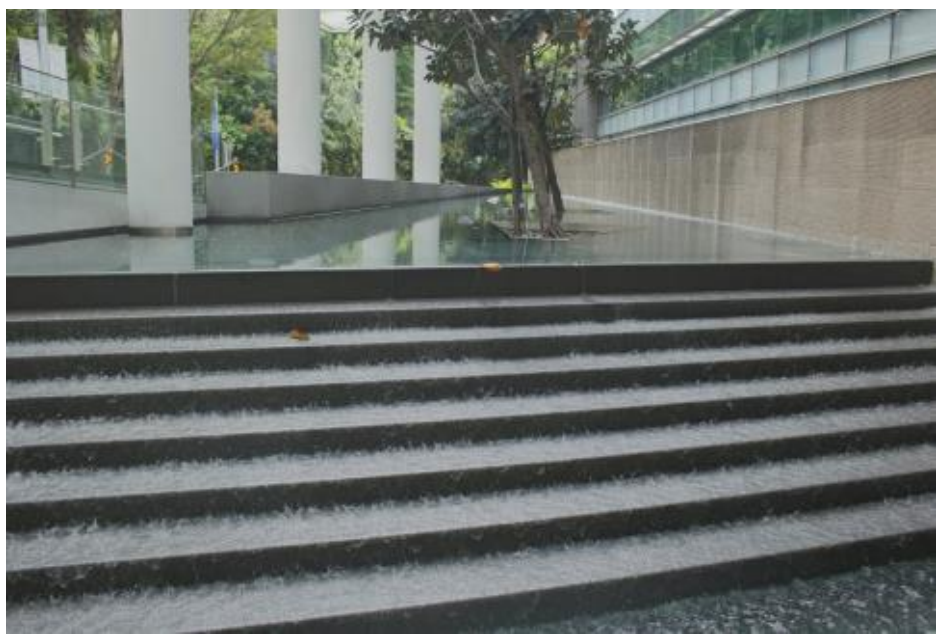

$V_{content}$ = 0 + 0.25 · 0.75 + 0.25 · 1 + 0.25· 1 = 0.688 (**Good**)

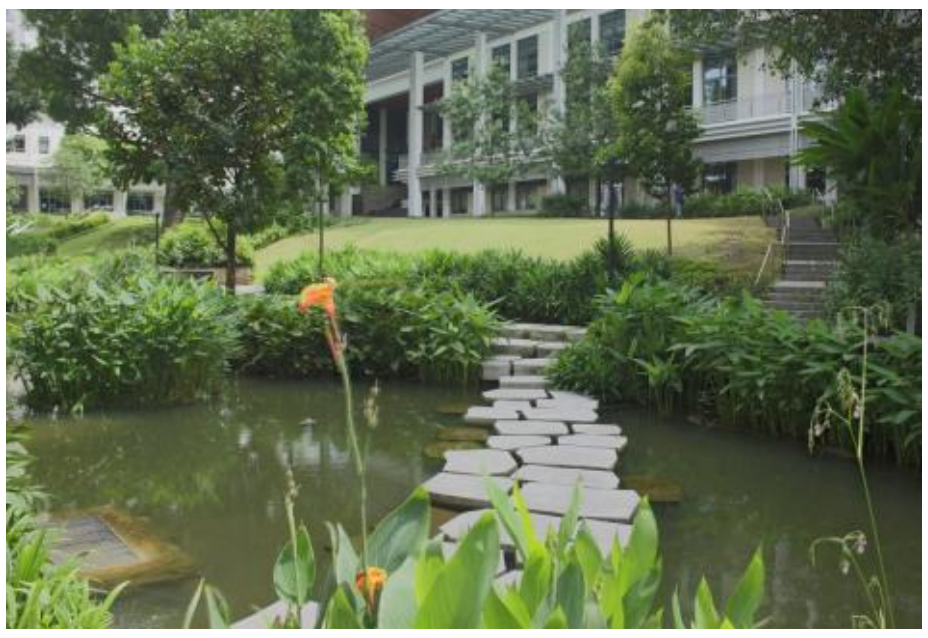

| | |
|---|---|
| $L_{sky}$ | 0 (no sky layer) |
| $L_{landscape}$ | 0.25 |
| $wf_{ct.dis.}$ | 1 (OLD 54) |
| $L_{ground}$ | 0.25 |
| $wf_{movement}$ | 1 (regular movement) |
| $L_{nature}$ | 0.25 |
| $wf_{nature}$ | 1 (> 50 %) |

$V_{content}$ = 0 + 0.25 · 1 + 0.25 · 1 + 0.25 · 1 = 0.75 (**Excellent**)

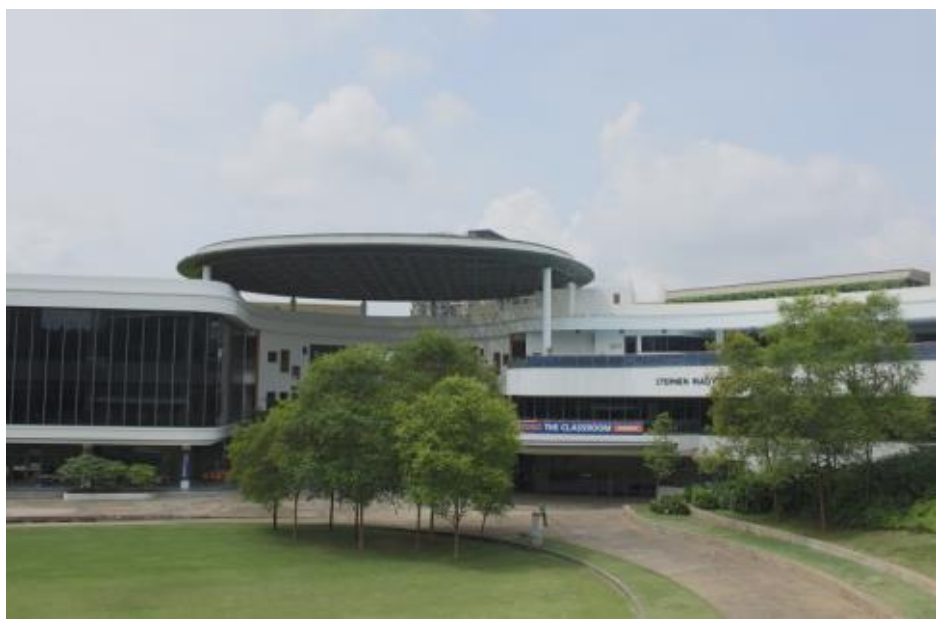

| | |
|---|---|
| $L_{sky}$ | 0.25 |
| $L_{landscape}$ | 0.25 |
| $wf_{ct.dis.}$ | OLD 63: 1 |
| $L_{ground}$ | 0.25 |
| $wf_{movement}$ | 0.5 (no regular movement) |
| $L_{nature}$ | 0.25 |
| $wf_{nature}$ | 0.75 (> 25 %) |

$V_{content}$= 0.25 + 0.25 · 1 + 0.25 · 0.5 + 0.25 · 0.75 = 0.813 (**Excellent**)

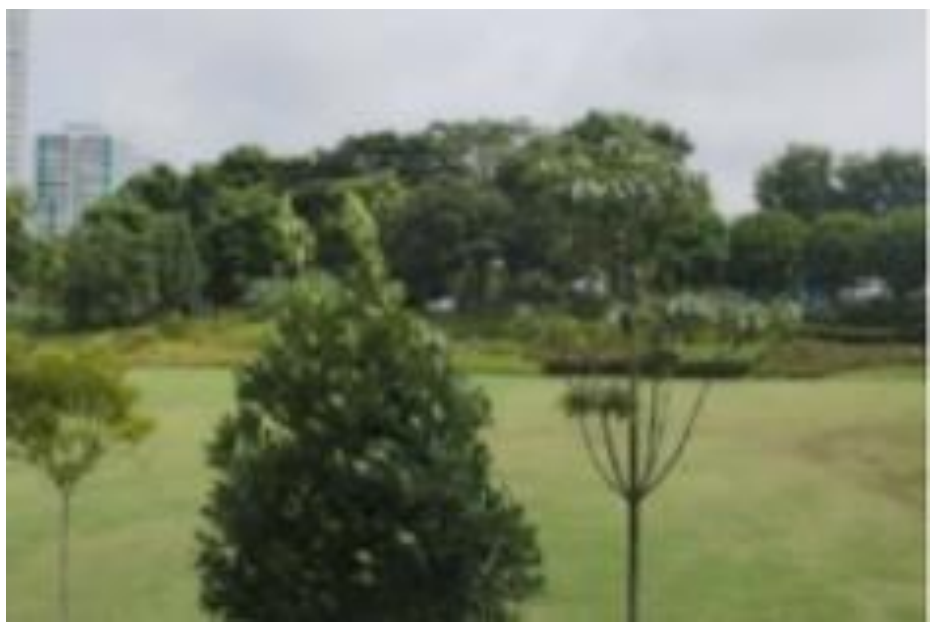

| | |
|---|---|
| $L_{sky}$ | 0.25 |
| $L_{landscape}$ | 0.25 |
| $wf_{ct.dis.}$ | 1 (OLD 65) |
| $L_{ground}$ | 0.25 |
| $wf_{movement}$ | 0.5 (no regular movement) |
| $L_{nature}$ | 0.25 |
| $wf_{nature}$ | 1 (> 50 %) |

$V_{content}$ = 0.25 + 0.25 · 1 + 0.25 · 0.5 + 0.25 · 1 = 0.875 (**Excellent**)

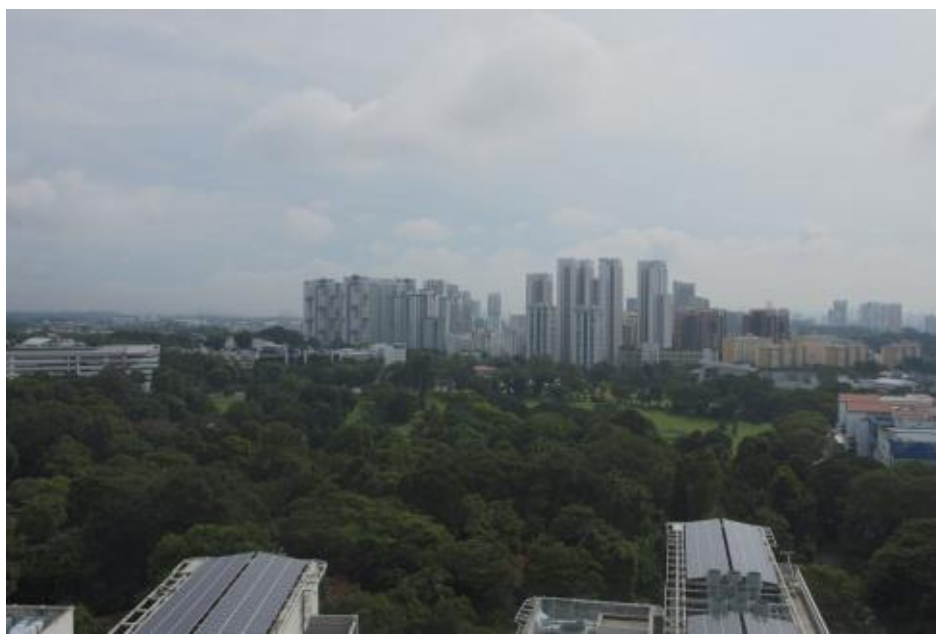

| | |
|---|---|
| $L_{sky}$ | 0.25 |
| $L_{landscape}$ | 0.25 |
| $wf_{ct.dis.}$ | 1 (OLD 851) |
| $L_{ground}$ | 0.25 |
| $wf_{movement}$ | 0.5 (no regular movement) |
| $L_{nature}$ | 0.25 |
| $wf_{nature}$ | 1 (> 50 %) |

$V_{content}$ = 0.25 + 0.25 · 1 + 0.25 · 0.5 + 0.25 · 1 = 0.875 (**Excellent**)

## Appendix G

In this section, we summarize future studies stemming from our literature review. Our framework is based on current design standards and scientific literature that examined different facets of view quality. Although this information is vital to both our work and the general field, their utility in the development of a design framework is often limited due to the narrow scope of the original work (particularly scientific literature). This generates many gaps that need to be filled by future research. To this, we have put forward a new research agenda for view assessment (Table 8). This summarizes key areas of research and assigns priority levels describing the urgency of each avenue of study.

**Table 8**. Summary of research needs

| Parameter | Primary variable responsible | Research questions | Priority |
|---|---|---|---|
| Variable weighting | All | • Do weightings vary according to the function of the space?<br>• Which view variable(s) better support the space's needs? | High |
| Minimum (acceptable) thresholds | | • What are the minimum characteristics of content, access, and clarity that produce quality views?<br>• What quantitative system best represents these thresholds? | |
| Dynamic features (movement) | Content | • How can we systematically evaluate the dynamic features in a view?<br>• Based on differences in scale (i.e., magnitude, frequency, and/or duration), what dynamic features need be considered (e.g., small birds, seasonal tree foliage, passing vehicles)? | |
| Composition of each view layer | | • How much of each layer (sky, landscape, ground) should be present in the view?<br>• Are all layers equally important, or should priority be given to one when designers have difficulty providing the other two? | Medium |
| Pleasant manmade elements | | How are aesthetically pleasing manmade elements (e.g., art, architecture, landmarks, etc.) perceived relative to other view content (e.g., nature)? | Low |
| Physical qualities | | Can the physical qualities of window view (e.g., spatial frequency) be used to measure its visual content, and also predict occupant responses? | Low |
| Multiple windows | Access | • How can we evaluate spaces (e.g., open-plan office) that have multiple windows or fragmented facades (e.g., clerestories)?<br>• When windows closely neighbor each other, can they be considered as one entity? | High |
| Window size and view angle | | When windows are measured using their physical dimensions (i.e., length and width) or by its solid angle, which provides a better indication of perceived view quality. | |
| Viewing distance and direction | | Does the quality of the view depend on the viewing distance and direction of the occupant? | Medium |
| Shading systems and facade materials | Clarity | When considering a wide range of shading systems and glazing materials, what are their holistic impacts on view retention, daylight access, and general protection (i.e., from glare and overheating, and to ensure privacy)? | High |

| | | |
|---|---|---|
| Dynamic shading | • How can we evaluate view perception when considering the constant changes associated with dynamic shading?<br>• What are the minimum design thresholds for visual clarity (e.g. how clear does the view need to be and for how long)? | |
| Partial shading | What scenarios need to be considered (e.g. venetian blinds, partial coverage of fabric shade) and how do we evaluate their effects on view quality? | Medium |